\documentclass[preprint,12pt]{elsarticle}

\usepackage{amssymb}
\usepackage{amsmath,amsfonts}
\usepackage[caption=false,font=scriptsize,labelfont=sf,textfont=sf]{subfig} % font=normalsize
\usepackage{graphicx}
\usepackage{multirow}
\usepackage{booktabs}
\usepackage{threeparttable}
\usepackage{multicol}  % Add this in the preamble if not already
\usepackage{color, soul}
\journal{Speech Communication}

\begin{document}

\begin{frontmatter}



\title{Advancing Automatic Speech Recognition using Feature Fusion with Self-Supervised Learning Features: A case study on Fearless Steps Apollo Corpus}


\author[inst1]{Szu-Jui Chen}

\affiliation[inst1]{organization={Center for Robust Speech Systems\\ Erik Jonsson School
of Engineering \& Computer Science},%Department and Organization
            addressline={\\University of Texas at Dallas}, 
            city={Richardson},
            postcode={75080}, 
            state={TX},
            country={USA}}

\author[inst1]{John H.L. Hansen\fnref{myfootnote}}

\fntext[myfootnote]{This project was funded, in part, by NSF-CISE Award 2016725, and partially by the University of Texas at Dallas from the Distinguished University Chair in Telecommunications Engineering held by J. H.L. Hansen.}

\begin{abstract}
Using self-supervised learning (SSL) models has significantly improved performance for downstream speech tasks, surpassing the capabilities of traditional hand-crafted features. This study investigates the amalgamation of SSL models, with the aim to leverage both their individual strengths and refine extracted features to achieve improved speech recognition models for naturalistic scenarios. Our research investigates the massive naturalistic Fearless Steps (FS) APOLLO resource, with particular focus on the FS Challenge (FSC) Phase-4 corpus, providing the inaugural analysis of this dataset. Additionally, we incorporate the CHiME-6 dataset to evaluate performance across diverse naturalistic speech scenarios. While exploring previously proposed Feature Refinement Loss and fusion methods, we found these methods to be less effective on the FSC Phase-4 corpus. To address this, we introduce a novel deep cross-attention (DCA) fusion method, designed to elevate performance, especially for the FSC Phase-4 corpus. Our objective is to foster creation of superior FS APOLLO community resources, catering to the diverse needs of researchers across various disciplines. The proposed solution achieves an absolute +1.1\% improvement in WER, providing effective meta-data creation for the massive FS APOLLO community resource.
\end{abstract}


% \begin{highlights}
% \end{highlights}


\end{frontmatter}


\section{Introduction}
% We can say e2e systems are dominant now. Recently better systems are achieved by large models with lots of training data (whisper using fbank). Traditionally we use fbank, but with the emergence of SSLR, we achieve much better performance with these features.
% TODO: Add a section talk about why previous system won't work but ours will.
End-to-end (E2E) automatic speech recognition (ASR) systems have emerged as the dominant solution in the research domain, over traditional hybrid HMM-DNN systems due to their simple training procedure and greater performance improvements \cite{xiong2018microsoft, watanabe2017hybrid}. Several E2E ASR systems \cite{kim2023branchformer, tuske2021limit, radford2022robust} have achieved state-of-the-art results on common datasets such as LibriSpeech \cite{panayotov2015librispeech}, Switchboard, and CHiME-6 \cite{watanabe2020chime}. Most studies have solutions that use conventional spectral features, such as an 80-dim. log-magnitude Mel spectrogram representation as input for model training.

\begin{figure}[!t]
    \centering
    \resizebox{1.1\columnwidth}{!}{
    \includegraphics{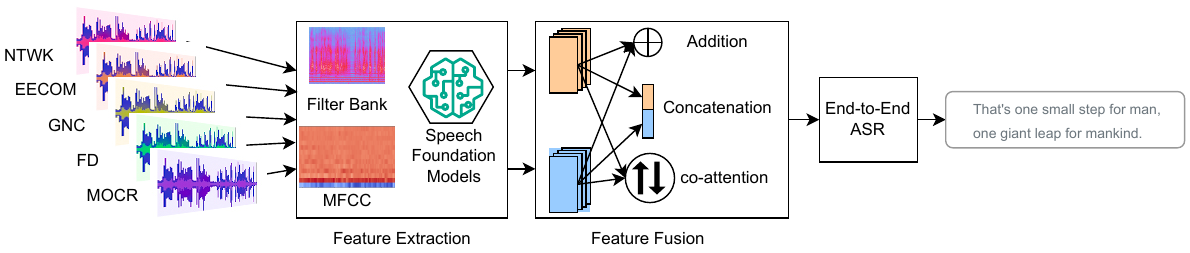}
    }
    \caption{Transcribing multi-channel naturalistic team-based audio using an end-to-end ASR system with feature fusion. Five of 30 parallel NASA Apollo communication loop channels are shown, all time synchronized with IRIG timecode Channel 1. The channels shown include Network Controller (NTWK), Electrical, Environmental, and Consumables Manager (EECOM), Guidance Navigation and Control (GNC), Flight Director (FD), and Mission Operations Control Room (MOCR).}
    \label{fig:general}
\end{figure}

% SSL features
In recent years, self-supervised learning (SSL) models have also shown remarkable performance benefits for a range of downstream tasks including speech translation \cite{nguyen2020investigating, wu2020self}, low resource ASR \cite{yi2020applying}, speaker verification, language ID \cite{fan2020exploring}, and emotion recognition \cite{pepino2021emotion}. These SSL models \cite{hsu2021hubert, baevski2020wav2vec, chen2022wavlm} leverage large amounts of unlabeled data during training, resulting in high-quality speech features that are well-suited for diverse downstream applications. One recent study \cite{chang2021exploration} demonstrated the superiority of using SSL representations (SSLR) over traditional hand-crafted features (e.g. FBANK). This leads to the basic question of whether combining a multiple set of SSL features would further enhance ASR systems. Several recent studies \cite{arunkumar2022investigation, chen2022fearless, chen2021scenario, berrebbi2022combining} have investigated the effectiveness of combining SSL features, or combining SSL and spectral features using alternate front-end and backend models along with fusion strategies. Fig.~\ref{fig:general} illustrates the general process of extracting features from multi-channel team-based speech signals, comparing traditional input features with SSL-based representations and showcasing fusion strategies such as addition, concatenation, and co-attention.

% talk about large ASR system and why they are not working in FSC.
Another promising path is training an ASR system using a vast and diverse range of audio data, resulting in enhanced resilience to accents, background noise, and specialized vocabulary/technical term content. One such example is the Whisper models developed by OpenAI \cite{radford2022robust}. These models exhibit strong generalization to standard benchmarks and frequently perform competitively with previous fully supervised results, all without requiring any fine-tuning in zero-shot transfer scenarios. Given these attributes, in this study we consider the Whisper model as a strong baseline in our evaluations.

This current study extends on our earlier preliminary investigation \cite{chen2022fearless} to further explore feature fusion with pre-trained SSL models. Our contributions can be highlighted as follows. First, we further investigate Feature Refinement Loss by exploring alternate parameter settings and then employing visualization tools to investigate the effect of the loss function. Next, a novel deep cross-attention (DCA) fusion solution is formulated based on SSL models and evaluated on the Fearless Steps Challenge (FSC) Phase-4 corpus, as well as CHiME-6 corpus. It will be shown that the proposed feature fusion method is effective when compared to other baseline methods, especially in naturalistic noisy speech scenarios. In addition, alternate state-of-the-art SSL models based on the SUPERB benchmark \cite{yang2021superb} are also explored. Building on these models, we present detailed phoneme-level error analysis, functional versus content word error analysis, and layer selection experiments to better understand the core strengths of fusion systems and the nature of performance improvements. Finally, this work is the first to present advanced ASR results as well as per-channel analysis of the FSC Phase-4 corpus, part of the extensive Fearless Steps APOLLO Community resource \cite{hansen2018fearless, hansen2024fearless}, comprising 150,000 hours of audio, meta-data, and speech technology infrastructure. While SSL models have demonstrated their potential across a wide range of speech processing tasks, this study focuses exclusively on ASR to evaluate and improve robustness of SSL models under challenging acoustic conditions. This targeted scope allows us to thoroughly explore fusion methods, such as our proposed DCA, and analyze their impact on ASR performance in real-world scenarios that include CHiME-6 and FSC Phase-4.

%%% General Figure %%%
%%%%%%%%%%%%%%%%%%%%

The study is organized as follows. We first discuss the relevant past studies in Sec.~\ref{related} and our proposed methods in Sec.~\ref{proposed}. Next, the experiment setups are presented in Sec.~\ref{setup}, with results and analyses in Sec.~\ref{results}. Finally, summary and conclusions are made in Sec.~\ref{conclusion}.
%%%% express in numbers %%%%
% The paper is organized as follows. We discuss the relevant past works in Sec.~2 %~\ref{related}
% and our proposed methods in Sec.~3 %~\ref{proposed}
% . Next, the experiment setups are presented in Sec.~4 %\ref{setup}
% and results and an ablation study are in Sec.~5. %\ref{results}
% Finally, conclusion are made in Sec.~6.%\ref{conclusion}.
%%%%%%%%%%%%%%%%%%%%%%%%%%%%

\section{Related Past Work}
\label{related}
\subsection{Self-Supervised Learning Representations}
% SSLR
Self-supervised learning (SSL) is a rapidly developing subset of unsupervised learning methods. These approaches leverage information derived from the input data itself to serve as the learning signal, enabling the acquisition of meaningful representations beneficial for subsequent tasks, especially for speech applications. SSL speech models can be divided into three main groups \cite{mohamed2022self}: generative approaches, contrastive approaches, and predictive approaches. We briefly introduce several of the most powerful SSL models in this section.\\

\textbf{Wav2Vec 2.0}: One of the top performing SSL models proposed recently is the Wav2Vec 2.0 \cite{baevski2020wav2vec}, which is categorized under contrastive approaches. By first passing the raw waveform through a convolutional feature encoder, applying an appropriate masking strategy, and subsequently utilizing a transformer network, we can have contextualized hidden representations $\mathbf{h}_t$ at time $t$. Next, a quantization module which uses a Gumbel softmax with a straight-through estimator are applied on these convolutional features, yielding quantized latent vectors. For each masked time step $t$, we define a set of quantized candidates $\mathbf{Q}_t = {\mathbf{q}_t, \tilde{\mathbf{q}}_1, \dots, \tilde{\mathbf{q}}_K}$, where $\mathbf{q}_t$ is the positive (true) quantized vector, and $\tilde{\mathbf{q}}_k$ for $k = 1, \dots, K$ are $K$ distractor vectors sampled uniformly from other masked time steps of the same utterance. Finally, the model employs the InfoNCE loss \cite{oord2018representation} to minimize the discrepancy between the contextualized hidden representations and the quantized vectors as follows:
\begin{equation}
\mathcal{L}_t = -\log\left(\frac{\exp(C(\mathbf{h}_t, \mathbf{q}_t) / K )}{\sum_{\tilde{\mathbf{q}} \sim \mathbf{Q}_t} \exp(C(\mathbf{h}_t, \tilde{\mathbf{q}}) / K )}\right),
\end{equation}
where $C(\cdot)$ is the cosine similarity function.\\ 

\textbf{HuBERT}: The Hidden Unit BERT (HuBERT) \cite{hsu2021hubert} approach, which is classified under predictive approaches, utilizes k-means units trained on MFCC features as the training target in the first iteration. In the subsequent iterations, it switches to using k-means units trained on latent representations. Similar to Wav2Vec 2.0, the HuBERT model also uses a convolutional feature encoder to take continuous waveform inputs and apply a certain masking strategy before the transformer network. The HuBERT model benefits from pre-computed k-means clusters as targets, allowing for a straightforward evaluation of the cross-entropy loss between the correct k-means cluster and the predicted cluster. In contrast, contrastive methods all require negative samples to prevent trivial solutions. The loss here is computed on both the masked $\mathcal{L}_m$ and unmasked $\mathcal{L}_u$ region, and is defined as:
\begin{align}
\begin{split}
\mathcal{L}_m = \sum_{t \in M} -\log p(z_t | X, t) ,\\
\text{and}\\
\mathcal{L}_u = \sum_{t \notin M} -\log p(z_t | X, t) ,\\
\end{split}
\end{align}
where $M$ is the total set of masked time steps and $z_t$ is the cluster unit. The final overall loss is calculated as $\mathcal{L} = \alpha \mathcal{L}_m + (1 - \alpha) \mathcal{L}_u$, where $\alpha$ denotes the weight between the two terms.\\

\textbf{WavLM}: WavLM \cite{chen2022wavlm} is similar to the HuBERT framework but equips the transformer with a gated relative position bias \cite{chi-etal-2022-xlm} to improve its capability on given recognition tasks. Different from other SSL models trained on single speaker data, WavLM introduces an utterance mixing strategy that enhances the training data by creating partially overlapped signals from different speakers to simulate more realistic mixed-type scenarios. In addition, during the pre-training phase, WavLM not only learns the masked speech prediction and denoising simultaneously, it also incorporates an extensive dataset of 94k hours of audio.

\subsection{Feature Fusion with SSLRs}
% Feature Fusion
Several studies have investigated the effectiveness of combining SSLRs for the downstream ASR model.
Early work by \cite{arunkumar2022investigation} examined the benefit of a combination of SSLRs from the last-layer outputs of the Wav2Vec 2.0 and HuBERT models, using a simple concatenation followed by a CTC layer on top. The models used are the pre-trained large variant that was fine-tuned with the LibriSpeech 960h dataset. Additionally, experiments were conducted by combining Wav2Vec 2.0, HuBERT, and WavLM models. The authors explored fine-tuning these models with either a Libri-100 hour dataset or the WSJ dataset, and showed the results for both base and large variant models. However, their use of simple concatenation and last-layer outputs limit the capacity to fully exploit complementary information across SSL models.
On the other hand, another study \cite{chen2022fearless} experimented with combining various SSL models, particularly their large variants if available, using different fusion methods for ASR. Instead of using features from the last layer, that study utilized multi-layer features extracted from the outputs of all layers in a pre-trained SSL model. These features were combined through a weighted-sum to form the final features for the downstream task. The study also demonstrated that there are correlations between the extracted SSL features, and further proposed a Feature Refinement Loss approach to better combine the SSL features.
Similarly, \cite{chen2021scenario} demonstrated the effectiveness of combining the general non-semantic SSLR with traditional MFCC features to address an ASR task. Their model exhibited the ability to identify different audio scenes during recognition.
Next, \cite{berrebbi2022combining} further extended fusion strategies by combining traditional spectral features with SSL features to address low resource datasets in both ASR and Speech Translation tasks. Here, a few learnable fusion methods were proposed, which included the co-attention based fusion and mixture of experts.

More recent works have introduced novel fusion mechanisms with improved scalability and performance.
For example, EFFUSE \cite{srivastava2024effuse} employs a distillation-based approach to train a single SSL model to predict the representations of multiple SSL models, achieving a +6.3\% average improvement on the Multilingual Speech Universal PERformance Benchmark (ML-SUPERB) while reducing parameter size by nearly half.
\textit{Wang et al.} \cite{wang2024fusion} introduced a fusion mechanism that integrates two discrete representations to enhance multilingual ASR performance. This approach preserves the benefits of discrete representations, such as reduced transmission and storage costs, while improving performance by integrating complementary information. They also explored self-augmented discrete representations, applying transformations to a single continuous SSL representation, thereby reducing inference costs. Experimental results on benchmarks like LibriSpeech and ML-SUPERB indicate up to 24\% relative character error rate improvement compared to non-fusion baselines.
Finally, \cite{chiu2024learnable} investigated methods for fusing feature representations derived from multiple speech SSL models, along with techniques to determine the optimal layer within each model. They evaluated five fusion strategies and found that temporal interleaved concatenation was the most effective. Additionally, they demonstrated that Gumbel layer selection can automatically select the most appropriate SSL layer, leading to better overall performance.

\section{Proposed Method}
\label{proposed}
In this section, we present our proposed fusion methods for combining features from SSL models. The combined features are then fed into a pre-encoder before the downstream encoder-decoder ASR model. We begin with an analysis of the Feature Refinement Loss, focusing on the impact of its hyper-parameters.

% Fig.~\ref{fig:frontend} provides an overall detailed flow diagram of deep cross-attention with two SSL models.

% Refinement loss
\subsection{Hyperparameter Analysis of Feature Refinement Loss}
\label{FR_loss}
The Feature Refinement Loss (FRL) \cite{chen2022fearless} is introduced in order to minimize redundancy among the SSL features when they are combined as input features for downstream speech tasks. This is accomplished by reducing the cross-correlation between the extracted SSL features.

The process can be formulated as follows. By submitting a segment of speech into two distinct pre-trained SSL models, we obtain the extracted features $\mathbf{X} \in {\mathbb{R}^{T_{1}\times D_{1}}}$ and $\mathbf{Y} \in {\mathbb{R}^{T_{2}\times D_{2}}}$, where $T_{1}$ and $T_{2}$ represent the input feature lengths, and $D_{1}$ and $D_{2}$ denote the respective feature dimensions. These features are computed as a weighted-sum over all hidden layers of the SSL models, where the weights are learnable and the SSL model parameters remain frozen. Since $D_{1}$ and $D_{2}$ are typically too large as inputs to an ASR model, we apply an affine transformation to $\mathbf{X}$ and $\mathbf{Y}$ to project them into a lower dimensional space of size $D$. Additionally, if $T_{1}$ and $T_{2}$ differ in feature size due to different time strides for the SSL models, we downsample $T_{1}$ to match the length of $T_{2}$ when $T_{1} > T_{2}$, denoted as $T$, and vice versa if $T_{2} > T_{1}$. We denote the affine transformation and downsample operation as \textsc{Norm} in our study. At this point, we have $\tilde{\mathbf{X}} = \textsc{Norm}(\mathbf{X}) \in {\mathbb{R}^{T\times D}}$ and $\tilde{\mathbf{Y}} = \textsc{Norm}(\mathbf{Y}) \in {\mathbb{R}^{T\times D}}$ for calculating the cross-correlation matrix $C \in {\mathbb{R}^{D\times D}}$ as follows,
%{\rm I\! R^{D\times D}
\begin{equation}
    C = MVN(\tilde{\mathbf{X}})^{\top} \cdot MVN(\tilde{\mathbf{Y}}), \\
\end{equation}
where $MVN(\cdot)$ is the mean and variance normalization along time $T$. With this cross-correlation matrix $C$, we can next define the Feature Refinement Loss $\mathcal{L_{\rm refine}}$ as:
\begin{equation}
    \mathcal{L_{\rm refine}} \overset{\Delta}{=}
    \sum_{i=1}^{D}{\sum_{j=1}^{D}{
    \begin{cases}
        (C_{\rm ij})^{2} & \text{if} \ |C_{\rm ij}| > \epsilon \\
        0 & \text{otherwise}
    \end{cases}}}, \\
\end{equation}
where $\epsilon$ controls the \textbf{\textit{maximum}} value of the correlation between the extracted features. Finally, we calculate the final overall loss $\mathcal{L}$ by combining the ASR loss and Feature Refinement Loss with a scaling combination parameter $\lambda$:
\begin{equation}
        \mathcal{L} = \mathcal{L_{\rm asr}} + \lambda \cdot \mathcal{L_{\rm refine}}. \\
\end{equation}
Note that only the affine transformation is affected by the Feature Refinement Loss, while the entire architecture is impacted by the ASR loss.

%%% correlation distribution %%%
\begin{figure}[t]
    \hspace{-2.5mm}
    \subfloat[Before training]{
    \includegraphics[width=.5\linewidth]{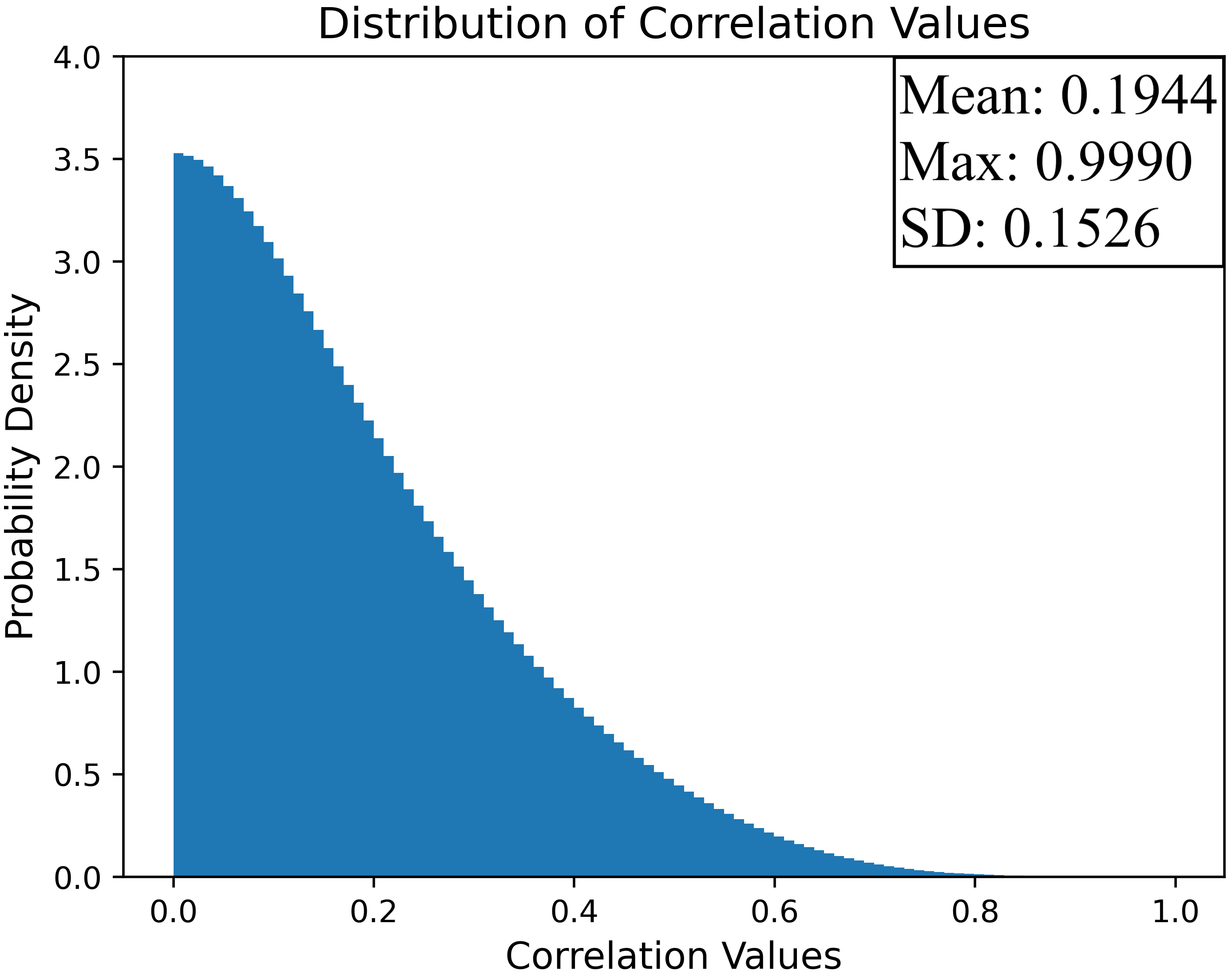}
    \label{fig:sub1} }
    \hspace*{\fill}
    \subfloat[After training]{
    \includegraphics[width=.5\linewidth]{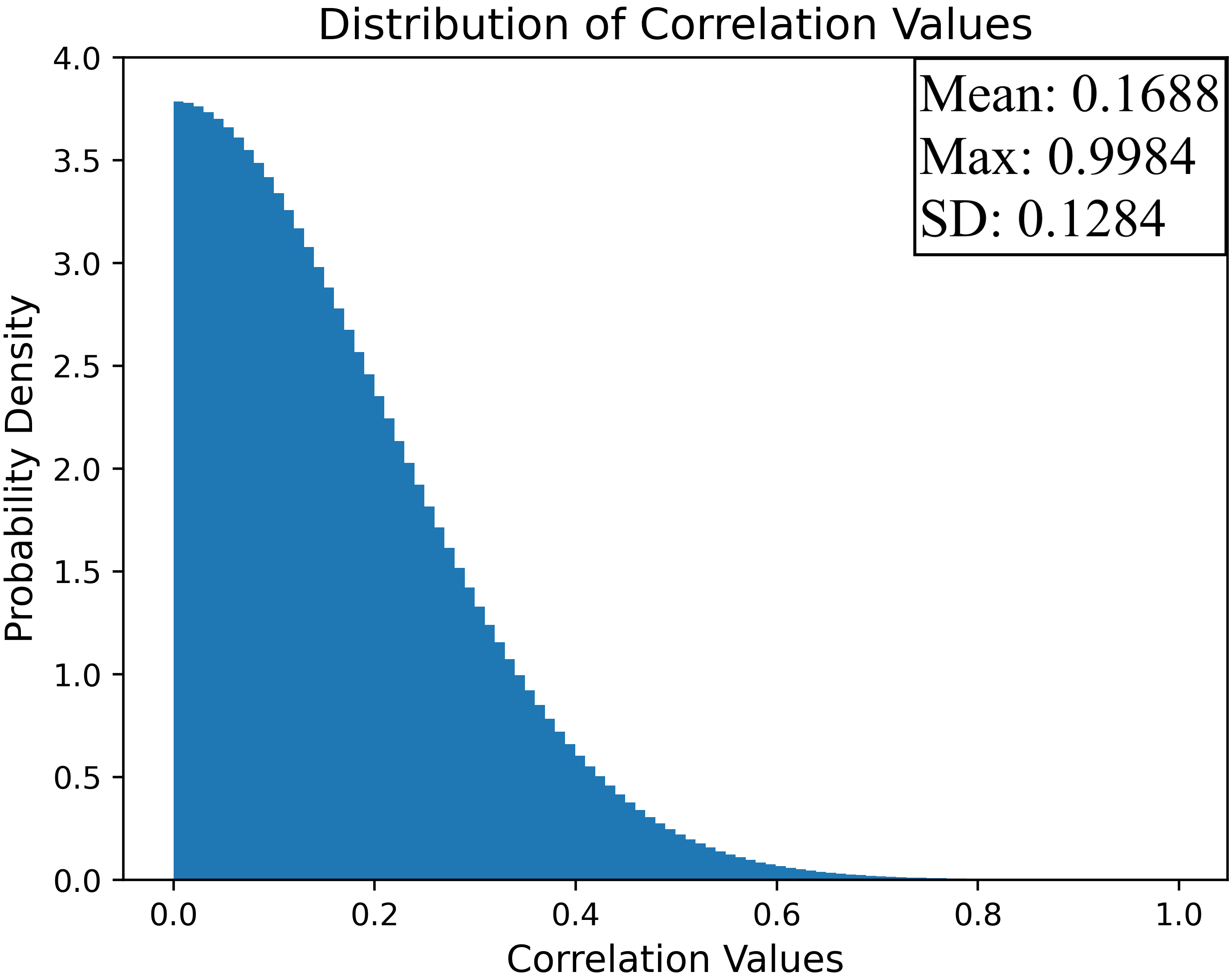}
    \label{fig:sub2} }
    \caption{Distribution of correlation values between HuBERT and Wav2Vec 2.0 features of entire FSC Phase-4 training set. (b) is the result when using $\epsilon=0.6$ and $\lambda=0.1$ for training.}
    \label{fig:corr_distribution}
\end{figure}
%%%%%%%%%%%%%%%%%%%%

As shown in Fig.~\ref{fig:corr_distribution}, we plot correlation distributions before and after training with HuBERT and Wav2Vec 2.0. From these plots, it is shown that we can force the absolute value of the correlation between HuBERT and Wav2Vec 2.0 models to be below 0.6, resulting in a better word error rate (WER) as will be shown in Table~4. %Table~\ref{tab:fusion}.

% Deep cross attention
\subsection{Deep Cross-Attention}
Recent advancements in cross-attention-based fusion have demonstrated its effectiveness across vision and natural language processing domains. In vision, \cite{lin2022cat} uses cross-attention to aggregate global information across feature maps, while \cite{chen2021crossvit} fuses multi-scale embeddings from large and small patch encoders. In NLP, \cite{cai2024car} leverages cross-attention to tackle cross-lingual summarization.

Building on these advancements, here we propose a deep cross-attention (DCA) method for fusing features extracted from SSL models. The core idea behind our method lies in leveraging the cross-attention mechanism to establish a meaningful connection between the layers of the two SSL models. Specifically, let $\mathbf{E}^{l}_{\text{A}} \in {\mathbb{R}^{T_{1}\times D_{1}}} | l \in \{1,\ldots,L_{1}\}$ and $\mathbf{E}^{m}_{\text{B}} \in {\mathbb{R}^{T_{2}\times D_{2}}} | m \in \{1,\ldots,L_{2}\}$ be the embeddings extracted from the $L_{1}$-layer SSL model A and $L_{2}$-layer SSL model B. By applying a weighted-sum on $\mathbf{E}^{l}_{\text{A}}$ and $\mathbf{E}^{m}_{\text{B}}$, we can now obtain the extracted features $\mathbf{X}$ and $\mathbf{Y}$ as formulated in Sec.~\ref{FR_loss}.
In this proposed DCA formulation, we incorporate two cross-attention modules for each of the layers, each consisting of a single-head scaled dot-product attention operation. This design choice is supported by preliminary experiments that showed using multi-head attention (e.g., 4 heads) led to worse performance than a single head.\footnote{In preliminary experiments, using 4 heads resulted in a 0.2\% higher WER.} To illustrate this, consider the output embedding from the first layer of the two models, denoted as $\mathbf{E}_{i} = \mathbf{E}^{l}_{i}$, with $l=1$. The query, key, and value vectors for the two cross-attention modules of the first layer can now be defined as:
\begin{equation}
\mathbf{Q}_i = \mathbf{E}_i \cdot \mathbf{W}^Q_i, \quad \mathbf{K}_i = \mathbf{E}_i \cdot \mathbf{W}^K_i, \quad \mathbf{V}_i = \mathbf{E}_i \cdot \mathbf{W}^V_i,
\end{equation}
where $i\in \{\text{A, B}\}$ and $\mathbf{W}^Q_{i}, \mathbf{W}^K_{i}, \mathbf{W}^V_{i}$ are learnable weight matrices of $\mathbb{R}^{D_{1}\times d_{\text{att}}}$ for the ``A2B'' attention module and of $\mathbb{R}^{D_{2}\times d_{\text{att}}}$ for the ``B2A'' attention module. Since the hidden embedding size of SSL models is often very large, we opt to use a lower dimension $d_{\text{att}}$ for the query, key, and value vectors. This helps manage the computational complexity while maintaining the overall effectiveness of the proposed method. Next, we can calculate the A2B-attended feature $\mathbf{E}_{\text{A2B}}$ and B2A-attended feature $\mathbf{E}_{\text{B2A}}$ as follow,
\begin{align}
\begin{split}
\mathbf{E}_{\text{A2B}} = \text{Softmax}\left(\frac{\mathbf{Q}_\text{A} \cdot \mathbf{K}_\text{B}^\top}{\sqrt{d_{\text{att}}}}\right) \cdot \mathbf{V}_\text{B},\\
\text{and}\\
\mathbf{E}_{\text{B2A}} = \text{Softmax}\left(\frac{\mathbf{Q}_\text{B} \cdot \mathbf{K}_\text{A}^\top}{\sqrt{d_{\text{att}}}}\right) \cdot \mathbf{V}_\text{A},\\
\end{split}
\end{align}
where the $(\cdot)$ is the dot product operation.

We can apply this methodology to each corresponding layer between model A and B. However, in cases where $L_1 \ne L_2$, we address the mismatch in depth by uniformly mapping the layers between the models. Without loss of generality, assume $L_1 < L_2$. For the A2B direction, we divide the $L_2$ layers of model B into $L_1$ consecutive segments. Each segment corresponds to a layer in model A, and we average the embeddings within each segment to form a mapped embedding $\tilde{\mathbf{E}}^l_{\text{B}}$ for the $l$-th layer. More formally, the $l$-th segment of model B covers layers from index $m_{\text{start}} = \left\lfloor \frac{(l-1) \cdot L_2}{L_1} \right\rfloor + 1$ to $m_{\text{end}} = \left\lfloor \frac{l \cdot L_2}{L_1} \right\rfloor,$
and we compute:
\[
\tilde{\mathbf{E}}^l_{\text{B}} = \frac{1}{m_{\text{end}} - m_{\text{start}} + 1} \sum_{m=m_{\text{start}}}^{m_{\text{end}}} \mathbf{E}^m_{\text{B}}.
\]
For the B2A direction, each layer $m \in \{1, \ldots, L_2\}$ in model B is assigned to a layer $l \in \{1, \ldots, L_1\}$ in model A using the rule:
\[
l = \left\lfloor \frac{(m - 1) \cdot L_1}{L_2} \right\rfloor + 1.
\]
Using these mappings, we compute the per-layer attended features $\mathbf{E}^l_{\text{A2B}}$ and $\mathbf{E}^m_{\text{B2A}}$, respectively. By applying a weighted-sum over these features, we obtain the cross-attended representations $\mathbf{F}_{\text{A2B}} \in \mathbb{R}^{T_{1} \times d_{\text{att}}}$ and $\mathbf{F}_{\text{B2A}} \in \mathbb{R}^{T_{2} \times d_{\text{att}}}$.
%%% all-pairs comparison %%%
% We can apply this methodology to each layer between model A and B. As such, with the weighted-sum operation on $\mathbf{E}^{l \rightarrow m}_{\text{A2B}}$ and $\mathbf{E}^{m \rightarrow l}_{\text{B2A}}$ for all $l \in \{1, \ldots, L_1\}$ in model A and $m \in \{1, \ldots, L_2\}$ in model B, we will obtain the cross-attended features $\mathbf{F}_{\text{A2B}} \in \mathbb{R}^{T_{1} \times d_{\text{att}}}$ and $\mathbf{F}_{\text{B2A}} \in \mathbb{R}^{T_{2} \times d_{\text{att}}}$ as defined below:
% \begin{equation}
% \begin{aligned}
% \mathbf{F}_{\text{A2B}} &= \sum_{l=1}^{L_1} \sum_{m=1}^{L_2} \alpha^{l \rightarrow m}_{\text{A2B}} \cdot \mathbf{E}^{l \rightarrow m}_{\text{A2B}}, \\
% \text{and}\\
% \mathbf{F}_{\text{B2A}} &= \sum_{m=1}^{L_2} \sum_{l=1}^{L_1} \alpha^{m \rightarrow l}_{\text{B2A}} \cdot \mathbf{E}^{m \rightarrow l}_{\text{B2A}},
% \end{aligned}
% \end{equation}
% where $\alpha^{l \rightarrow m}_{\text{A2B}}$ and $\alpha^{m \rightarrow l}_{\text{B2A}}$ are trainable weights normalized via softmax.
%%%%%%%%

The final feature therefore, denoted as $\mathbf{F}_{\text{ASR}}$, is a combination of the four extracted features $\mathbf{X}$, $\mathbf{Y}$, $\mathbf{F}_{\text{A2B}}$, and $\mathbf{F}_{\text{B2A}}$, written as:
\begin{equation}
    \label{eq8}
    \mathbf{F}_{\text{ASR}}=[\textsc{Norm}(\mathbf{X};\mathbf{F}_{\text{A2B}});\textsc{Norm}(\mathbf{Y};\mathbf{F}_{\text{B2A}})],
\end{equation}
where the semicolon represents the feature vector concatenation operation, and the \textsc{Norm} denotes the affine transformation and downsampling, resulting in the feature representation $\mathbf{F}_\text{ASR} \in {\mathbb{R}^{T\times 2D}}$. Here, the affine transformation reduces the dimensionality of each feature to $D$, and the final $\mathbf{F}_{\text{ASR}}$ feature has a dimension of $2D$, since it combines two such transformed features.
%Note that the final feature dimension $D$ can be either $D_{1}+D_{att}$ or $D_{2}+D_{att}$, where $D_{1}$ and $D_{2}$ are typically the same.

In our early experiments, we have investigated different ways of combining these four features, including (1) concatenating normalized and cross-attended features, $[[\textsc{Norm}(\mathbf{X});\mathbf{F}_{\text{A2B}}];[\textsc{Norm}(\mathbf{Y});\mathbf{F}_{\text{B2A}}]]$; (2) summing the normalized and cross-attended features $[\textsc{Norm}(\mathbf{X}) + \mathbf{F}_{\text{A2B}}];[\textsc{Norm}(\mathbf{Y}) + \mathbf{F}_{\text{B2A}}]$; (3) weighted-sum of $\textsc{Norm}(\mathbf{X}; \mathbf{F}_{\text{A2B}}) \text{ and } \textsc{Norm}(\mathbf{Y}; \mathbf{F}_{\text{B2A}})$; and (4) using only the cross-attended features, $[\mathbf{F}_{\text{A2B}};\mathbf{F}_{\text{B2A}}]$. However, none of these alternatives proved to be more effective than our proposed solution shown in Eq.~\ref{eq8}.\footnote{WER increases for the four feature set combinations highlighted as options (1)-(4), with absolute increases of 0.2\%, 0.1\%, 0.3\%, and 4.9\%, respectively.}
% Dev set: (1)25.4, (2)25.4, (4)29.4, eq8:24.9
% Eval set: (1)27.5, (2)27.3, (4)32.2, eq8:27.3

%%% Frontend Model %%%
\begin{figure*}[!t]
    \centering
    \scalebox{.8}{
    \includegraphics{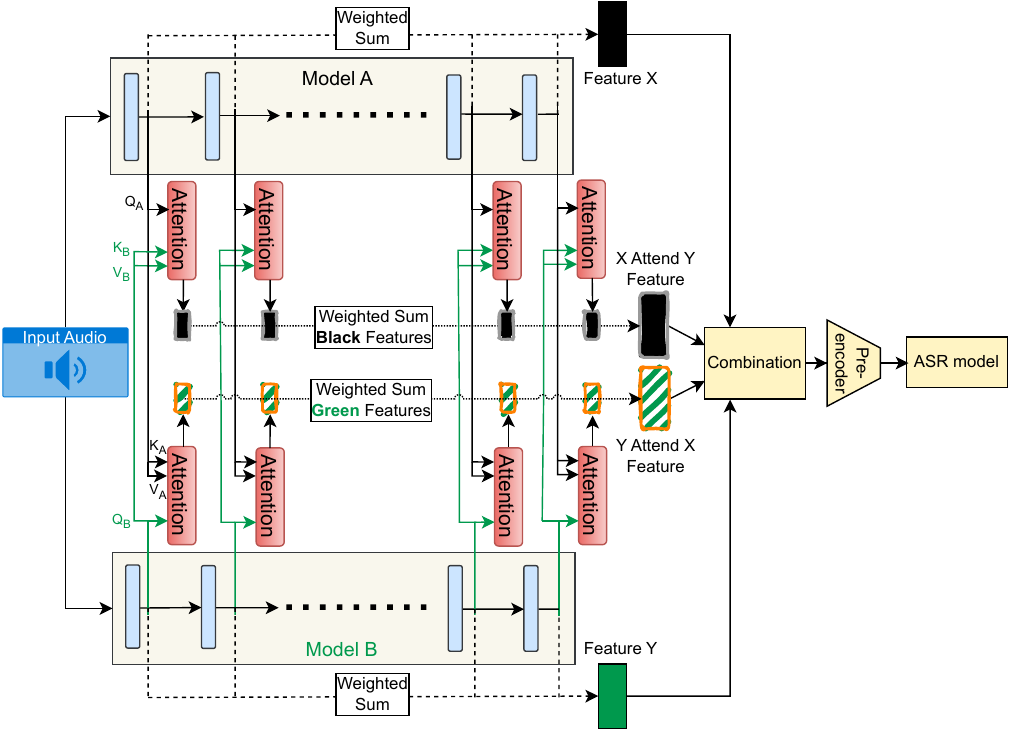}
    }
    \caption{The deep cross-attention feature fusion with two self-supervised learning models. The figure shows how the output of each layer is used to attend to the corresponding layer, generating X-Attend-Y and Y-Attend-X features (i.e., $\mathbf{F}_{\text{A2B}}$ and $\mathbf{F}_{\text{B2A}}$ in Eq.~\ref{eq8}) as extra inputs.}
    \label{fig:frontend}
\end{figure*}
%%%%%%%%%%%%%%%%%%%%
The philosophy of the proposed DCA fusion is to capture the interplay between the representations learned by the two models, facilitating the extraction of complementary and discriminative speech context features.
As shown in Fig.~\ref{fig:frontend}, we can use the output representations from each layer for the cross-attention operation. 

\section{Experimental Setup}
\label{setup}
% Dataset
\subsection{Dataset}
\subsubsection{Fearless Steps Challenge Corpus}
\label{FSC_intro}
The Fearless Steps Challenge (FSC) corpus is a subset of the original 19,000-hour (2017) Fearless Steps Corpus \cite{hansen2018fearless}\footnote{The Fearless Steps APOLLO community Resource, under NSF support, continues to expand and will encompass +150,000 hours of naturalistic team communications from the NASA Apollo missions, Gemini, and sample ISS-International Space Station, along with broadcast Public Affairs Officer (PAO) news data (see: exploreapollo.org)}. The FSC portion encompasses Phases 1 through 4 \cite{Hansen2019, joglekar2020fearless, Joglekar2021}, with prior research predominantly concentrated on the FSC Phase-2 dataset \cite{gorin400houston, chen2021scenario,  chen2022fearless}. The FSC Phase-1 and Phase-2 corpora comprise 40 and 100 hours of labeled data, respectively, from five team active channels of Apollo-11: Network Controller (NTWK), Electrical, Environmental and Consumables Manager (EECOM), Guidance, Navigation, and Control (GNC), Flight Director (FD), and Mission Operations Control Room (MOCR). Later, an additional 9 hours of data from a previously unseen Apollo-11 channel Operations and Procedures (OPSPRO) and 5 hours of data from Apollo-13 were incorporated into the Phase-3 corpus for open unseen testing. Subsequently, FSC Phase-4 further extended the corpus by introducing an additional 6 hours of Apollo-8 data, resulting in a total of 120 hours of full transcribed/meta-data audio material. A summary of FSC corpus Phases is shown in Table~\ref{tab:fsc_summary}. It is important to note that both training and development sets of FSC Phase-4 corpus exclusively contain data from the original five selected channels of Apollo-11. Consequently, FSC Phase-4 presents a significantly more challenging dataset compared to previous versions, due to the inclusion of unseen channel speakers/loops/conditions and missions. In our study, we focus on the ASR Track-2 of FSC Phase-4 corpus, which contains segmented audios. To summarize, the segmented data comprises a total of 29.8 hours for training, 8.6 hours for development (Dev), and 19.2 hours for evaluation (Eval).

\begin{table}[ht]
\centering
\caption{Summary of FSC corpus Phases. Column 2 shows the hours of labeled data provided for train, dev, and eval sets.}
\scalebox{0.9}{
\begin{tabular}{c|c|p{2cm}|p{7.5cm}}
\toprule
\textbf{Phase} & \textbf{Size} & \textbf{Mission} & \textbf{Distinguishing Characteristics} \\
\hline
1 & 0/20/20 hrs & Apollo-11 & Unsupervised/semi-supervised system development for the five team channels: NTWK, EECOM, GNC, FD, and MOCR. \\
\hline
2 & 65/15/20 hrs & Apollo-11 & Same five channels as Phase-1. Most prior research is based on this phase. \\
\hline
3 & 65/15/34 hrs & Apollo-11, Apollo-13 & Adds 9 hrs from unseen channel (OPSPRO) of Apollo-11 and 5 hrs from Apollo-13 to eval set. \\
\hline
4 & 67/18/35 hrs & Apollo-8, Apollo-11, Apollo-13 & Adds 6 hrs of Apollo-8 data. Training/dev sets include only the original five Apollo-11 channels. \\
\bottomrule
\end{tabular}
}%scaling
\label{tab:fsc_summary}
\end{table}

\subsubsection{CHiME-6}
The CHiME-6 corpus \cite{watanabe2020chime} is a challenging dataset designed for ASR in real-world, multi-speaker environments. It consists of conversational speech recorded during dinner parties in domestic settings, featuring overlapping speech, background noise, and microphone variability. These characteristics make CHiME-6 a valuable resource for evaluating the robustness of ASR systems under naturalistic and adverse acoustic conditions. We follow the CHiME-6 recipe in ESPnet \cite{watanabe2018espnet}, which uses guided source separation \cite{boeddeker2018front} to enhance dev/evaluation sets. However, we do not apply the speed perturbation and language model for CHiME-6 experiments.

% \subsubsection{LibriSpeech}
% We also conduct experiments on the LibriSpeech corpus \cite{panayotov2015librispeech}. Our models are trained on the 100-hour subset (train-clean-100) and evaluated on the mainstream test sets: dev-clean/other and test-clean/other. 

% System
% \subsection{Backend ASR System}
\subsection{Model, Optimization and Evaluation}
All experiments are conducted using the ESPnet toolkit \cite{watanabe2018espnet}. For the FSC corpus, we evaluate both Conformer and E-Branchformer-based ASR models, while CHiME-6 corpus experiments focus exclusively on E-Branchformer. Details of the model architecture, optimization strategies, and evaluation methods are provided below. 
\begin{enumerate}
    \item \textit{Model}: For SSL models, we mainly consider the large versions of Wav2Vec 2.0, HuBERT, and WavLM. These front-end models are kept frozen during training, serving solely as feature extractors. As a result, we only report the number of trainable parameters in the tables. Between the front-end feature extractor and backend ASR model, we have a pre-encoder layer that converts the features into a 80-dimensional feature vector. The backend ASR model employs a hybrid CTC/Attention architecture \cite{watanabe2017hybrid}, using either a 12-layer Conformer encoder \cite{gulati2020conformer} or a 12-layer E-Branchformer encoder \cite{kim2023branchformer}, paired with a 6-layer Transformer decoder \cite{vaswani2017attention}. All attention modules in the Conformer encoder, E-Branchformer encoder, and Transformer decoder use an attention dimension of 256 and 4 attention heads. The Conformer encoder and Transformer decoder have a feed-forward dimension of 2048, whereas the E-Branchformer encoder uses a feed-forward dimension of 1024. The CNN module in each Conformer layer has 15 kernels, while E-Branchformer layers use 31 kernels. In the DCA fusion method, all attention modules use an attention dimension of $d_{\text{att}}=100$. For the affine transformation in the $\textsc{Norm}$ operation discussed in Sec.~\ref{proposed}, we set the feature dimension to $D=100$ for our study. Due to our GPU memory limitations, we apply the DCA operation only to the even layers of the SSL models.
    \item \textit{Optimization}:
    For experiments using the Conformer encoder on the FSC corpus, we adopted the Adam optimizer \cite{kingma2014adam} with a warmup learning rate scheduler, linearly increasing the learning rate to 0.001 over 25k steps, followed by exponential decay. For experiments using the E-Branchformer encoder, we employed the AdamW optimizer \cite{loshchilov2017decoupled}. On the FSC corpus, the learning rate was warmed up to 0.001 over 25k steps, whereas for the DCA fusion experiments, a learning rate of 0.002 was used with a shorter warmup phase of 15k steps. For the CHiME-6 corpus, the learning rate was warmed up to 0.001 over 20k steps, with 20k warmup steps also applied in the DCA fusion experiments. To maximize GPU usage, we employed ESPnet’s \texttt{numel} sampler with a batch bins size of 4 million. We consistently used the SpecAugment \cite{park2019specaugment} with two time masks and two frequency masks. All models are trained on 8 NVIDIA 2080Ti GPUs.
    \item \textit{Evaluation}: We use the average of checkpoints from the top ten epochs and top five epochs for FSC and CHiME-6 experiments respectively. During decoding, we applied a Transformer language model (LM) trained from the transcripts of the training set, with a weight of 0.1 for the FSC corpus. Also, a LM was not applied for the CHiME-6 experiments. Statistical significance is evaluated using the NIST SCTK Matched-Pair Sentence Segment Word Error (MAPSSWE) test \cite{gillick1989some, fiscus2018sctk}. MAPSSWE is applied on the full FSC Phase-4 Eval set (22,025 sentences) for FSC results, and on the CHiME-6 Eval set (11,027 sentences) for experiments conducted on CHiME-6.
\end{enumerate}
To summarize, the FSC corpus experiments include both Conformer-based models (Tables~\ref{tab:p2p4} and \ref{tab:tune_refine_loss}) and E-Branchformer-based models (Tables~\ref{tab:combine}-\ref{tab:wer_missions}), while the later CHiME-6 experiments exclusively use only the E-Branchformer architecture (Table~\ref{tab:chime}).

\section{Experimental Results and Analysis}
\label{results}
\subsection{From FSC Phase-2 to Phase-4}
\begin{table}[t]
\centering
\caption{WER (\%) performance comparison between Fearless Steps Challenge Phase-2 (5 channel loops) and Phase-4 (5 channel loops with added unseen channels and missions). We include state-of-the-art results on FSC Phase-2 corpus and performance of pre-trained ASR model by OpenAI for comparison (e.g., references \cite{radford2022robust,chen2021scenario,gorin400houston}). Please note that Phase-2 and Phase-4 results are not directly comparable.}
% \scalebox{0.75}{
% \textsc{LARGE} originally afte hubert and wavlm
\begin{threeparttable}
  \begin{tabular}{l|c|c|c}
    \toprule
    \textbf{Approaches} & \textbf{FSC Phase} & \textbf{Dev}$(\downarrow)$ & \textbf{Eval}$(\downarrow)$\\
    \hline
    Whisper medium~\cite{radford2022robust} zero-shot & \multirow{6}{*}{Phase-2} & 53.1 & 55.5\\
    Whisper base~\cite{radford2022robust} fine-tuned & & 29.3 & 27.6\\
    \textit{Gorin et al.} \cite{gorin400houston} & & 28.6 & 31.4\\
    \textit{Chen et al.} \cite{chen2021scenario} & & 26.2 & 28.9\\
    HuBERT + Enc./Dec. & & 31.8 & 34.2\\
    WavLM + Enc./Dec. & & \textbf{22.7} & \textbf{24.7}\\
    %\textit{Gorin et al.}~\cite{gorin400houston} best~\tnote{\textdagger} & 2 & 21.8 & 24.3\\
    \midrule
    HuBERT + Enc./Dec. & \multirow{2}{*}{Phase-4} & 37.0 & 38.6\\
    %\hline
    WavLM + Enc./Dec. & & \textbf{25.1} & \textbf{27.7}\\    
    \bottomrule
  \end{tabular}
% }
  % \begin{tablenotes}
  %   \item[\textdagger] Used FSC corpus plus 19k hours of unlabeled A11 corpus for semi-supervised training.
  %   %\item[$^\ddagger$] fine-tuned
  % \end{tablenotes}
\end{threeparttable}
  \label{tab:p2p4}
\end{table}
% FSC p2 vs p4 comparison (add Whisper result
% OLD: With the addition of 19k-hours of unlabeled Apollo 11 corpus data used for semi-supervised training, \textit{Gorin et al.} pushed their WER to 24.3\% on the Eval set.
%As all the past works on FSC corpus focus on Phase-2, we would like to show the performance of
This work is the first to report results on the FSC Phase-4 corpus, while all prior studies have evaluated only on Phase-2. To provide meaningful context for our Phase-4 results, we include comparisons with the same systems evaluated on both Phase-2 and Phase-4. First, we present results of previous works on FSC Phase-2 corpus as shown in Table~\ref{tab:p2p4}. The previous work by \textit{Chen et al.} \cite{chen2021scenario} achieved 28.9\% word error rate (WER) on the Eval set, outperforming the previous best result of 31.4\% by \textit{Gorin et al.} by 2.5\%. For our baseline, we also used the Whisper model by OpenAI for zero-shot inference and fine-tuning. For zero-shot inference, the medium variant of Whisper model that is with 769 million parameters gives a 55.5\% WER. When fine-tuning the base variant that is with 74 million parameters with FSC Phase-2 audio corpus, we observe a 27.6\% WER, which is better than all previous works using the same amount of training data for FSC.

Next, we present results of our baseline models tested on both FSC Phase-2 and Phase-4 to demonstrate the performance implications when dealing with unseen channels and missions scenarios present in Phase-4. When using extracted features from HuBERT or WavLM, we observe a -4.4\% and -3.0\% absolute WER degradation on Eval sets from FSC Phase-2 to Phase-4. This demonstrates how challenging it is to achieve effective ASR performance for models with unseen channels and unseen missions. Notably, the system that uses features extracted from WavLM achieves a 24.7\% WER on FSC Phase-2, outperforming both previous works and the Whisper-based models.

\begin{table}[t]
\centering
\caption{WER (\%) on FSC Phase-4 corpus when changing $\lambda$ and $\epsilon$ of Feature Refinement Loss: The system uses HuBERT and Wav2vec 2.0 as feature extractors, combined by linear projection method.}
\begin{threeparttable}
  \begin{tabular}{c|c|c|c|c||c|c}
    \toprule
    &&& \multicolumn{2}{c||}{with LM} & \multicolumn{2}{c}{without LM} \\
    \hline
    \textbf{Row\#} & \textbf{$\lambda$} & \textbf{$\epsilon$} & \textbf{Dev}$(\downarrow)$ & \textbf{Eval}$(\downarrow)$ & \textbf{Dev}$(\downarrow)$ & \textbf{Eval}$(\downarrow)$ \\
    \hline
    0 & - & - & 36.3 & 38.6 & 36.7 & 39.0\\
    \hline
    1 & 0.1 & 0.8 & 35.9 & \textbf{38.1} & 36.4 & 39.0\\
    2 & 0.1 & 0.6 & \textbf{35.6} & 38.2 & \textbf{35.9} & \textbf{38.5}\\
    3 & 0.1 & 0.4 & 36.3 & 38.4 & 36.7 & 39.0\\
    4 & 0.1 & 0.2 & 36.5 & 38.8 & 36.9 & 39.4\\
    5 & 0.1 & 0.1 & 36.9 & 39.7 & 37.4 & 40.1\\
    \hline
    6 & 0.1 & 0.8up\tnote{\textdagger} & 37.4 & 39.9 & 37.6 & 40.3\\
    7 & 0.1 & 0.6up\tnote{\textdagger} & 37.5 & 39.9 & 37.8 & 40.3\\
    8 & 0.1 & 0.4up\tnote{\textdagger} & 36.7 & 39.1 & 37.1 & 39.5\\
    9 & 0.1 & 0.2up\tnote{\textdagger} & 37.1 & 39.1 & 37.4 & 39.6\\
    \hline
    10 & 0.5 & 0.6 & 36.0 & 38.4 & 36.3 & 38.8\\
    11 & 0.01 & 0.6 & 35.9 & \textbf{38.1} & 36.1 & \textbf{38.5}\\
    12 & 0.005 & 0.6 & 35.8 & 38.3 & 36.5 & 38.7\\
    13 & 0.5 & 0.1 & 36.9 & 39.4 & 37.5 & 40.1\\
    14 & 0.01 & 0.1 & 36.2 & 38.4 & 36.8 & 38.8\\
    15 & 0.005 & 0.1 & 35.8 & 38.3 & 36.3 & 39.1\\
    \bottomrule
  \end{tabular}
  \begin{tablenotes}
    \item[\textdagger] These entries refer to experiments in which the model is trained with a \textbf{\textit{minimum}} $\epsilon$ correlation value.
  \end{tablenotes}
\end{threeparttable}
\label{tab:tune_refine_loss}
\end{table}
\subsection{Analysis of Feature Refinement Loss}
\label{FRL_result}
% loss parameter tuning on FSC p4
In this section, we investigate the hyper-parameters, $\epsilon$ and $\lambda$, employed in the Feature Refinement Loss (FRL). We combine the features from HuBERT and Wav2Vec 2.0 using the linear projection method described in \cite{chen2022fearless} as our baseline, shown in Row 0 of Table~\ref{tab:tune_refine_loss}. Since FRL primarily targets improvements in acoustic feature representations, we also report results without a language model (LM) to better assess the impact of FRL on the acoustic modeling without influence of LM decoding. We first explore the maximum value of correlation between the extracted features by varying $\epsilon$, as shown in Rows 1 to 5 of Table~\ref{tab:tune_refine_loss}. When using the LM, the best result is achieved when $\epsilon=0.6$ for the Dev set, and $\epsilon=0.8$ for Eval set. These results demonstrate that maintaining the maximum correlation between the extracted features from $\epsilon=0.4$ to $\epsilon=0.8$ improves WER. However, constraining the correlation too tightly ($\epsilon<0.4$) leads to performance degradation. The results without an LM show that only $\epsilon=0.6$ improves the WER over the baseline, suggesting that FRL primarily enhances the quality of acoustic representations when the correlation constraint is moderately relaxed. The improvement of best-performing configuration (Row 2) is statistically significant with $p<0.001$, as verified using the MAPSSWE test \cite{gillick1989some}.

Next, we investigate the effect of increasing the correlation between the extracted features in Rows 6 to 9. Unfortunately, this direction produces worse WER either with or without a LM. Therefore, it is suggested that strengthening the correlation amongst features from SSL models is not a positive research path.

Lastly, we investigate the scaling combination parameter $\lambda$ for leveraging the Feature Refinement Loss (Rows 10–15). When $\epsilon = 0.6$, improvements are observed for $\lambda$ from 0.005 to 0.5. In contrast, for $\epsilon = 0.1$, smaller $\lambda$ values perform better, reinforcing the idea that overly constraining correlations harms performance unless the constraint is applied very lightly.
When $\lambda=0.005$, both $\epsilon=0.6$ (Row 12) and $\epsilon=0.1$ (Row 15) achieve WERs comparable to the best-performing configuration at $\lambda=0.1$ (Rows 1 and 2). However, unlike at $\lambda=0.1$ where performance varies clearly with different $\epsilon$ values, the WERs at $\lambda=0.005$ remain relatively flat across a wide $\epsilon$ range (Rows 12 and 15 vs. Rows 10 and 13). This suggests that a small $\lambda$ may suppress the impact of $\epsilon$, making the model less sensitive to the intended correlation constraints. In other words, while FRL still brings benefits at $\lambda=0.005$, the interaction between $\lambda$ and $\epsilon$ appears weaker, reducing the effectiveness of fine-tuning $\epsilon$ at very low $\lambda$ values.
%When $\lambda=0.005$, either $\epsilon=0.6$ or $0.1$ gives the same results, which implies the value of $\lambda$ might be too small for $\epsilon$ to be effective at this point.
As a result, we use $\lambda=0.1$ and $\epsilon=0.6$ for all of the remaining experiments.

In conclusion, our results suggest that the Feature Refinement Loss is most effective when the correlation threshold $\epsilon$ is set to a moderate value (0.6) and paired with a sufficiently strong scaling parameter ($\lambda=0.1$). In contrast, low $\epsilon$ values require weaker regularization (smaller $\lambda$), and pushing correlations to be too small or too large leads to suboptimal results. These insights highlight the importance of carefully tuning both parameters to balance constraint strength with learning flexibility.
%In conclusion, we observe that the Feature Refinement Loss benefits from either a smaller weight ($\lambda<0.01$) with low correlation ($\epsilon=0.1$) or a larger weight ($\lambda=0.1$) with moderate correlation ($\epsilon=0.6$). Specifically, for low correlation ($\epsilon=0.1$), smaller $\lambda$ values (e.g., 0.005) outperform larger ones, suggesting that imposing too strong a constraint on the feature correlation harms performance. In contrast, for moderate correlation ($\epsilon=0.6$), larger $\lambda$ values (e.g., 0.1) are more effective, as they allow the Feature Refinement Loss to operate more effectively without overly restricting the feature interactions. This highlights the importance of carefully tuning both $\lambda$ and $\epsilon$ to achieve optimal results from the Feature Refinement Loss.

\subsection{Result of Combining SSLR}
\label{sec:combine_ssl}
%%% E-branchformer updated %%%
\begin{table}[t]
\centering
\caption{WER (\%) on FSC Phase-4 corpus when combining WavLM with different SSL models and Fbank feature using linear projection method. The second column reports \textit{total} number of model parameters in millions (M), including frozen SSL models, and the third column shows number of \textit{trainable} parameters. The second last column reports the substitution (S), deletion (D), and insertion (I) error rates (\%) from the Eval set. The last column shows the relative contribution of each SSL model (based on projection weight norms), corresponding to the order of models in the first column. E-Branchformer is used in these table results.}
\scalebox{0.65}{
  \begin{tabular}{l|c|c|c|c|c|c}
    \toprule
    \textbf{SSL Models} & \textbf{Total(M)} & \textbf{Trainable(M)} & \textbf{Dev}$(\downarrow)$ & \textbf{Eval}$(\downarrow)$ & \textbf{S$|$D$|$I} & \textbf{Weight(\%)}\\
    \hline
    Data2Vec (D2V) & 349.6 & 36.3 & 38.1 & 37.9 & $23.5|8.1|6.3$ & -\\
    HuBERT (HB) & 352.9 & 36.3 & 35.0 & 36.7 & $22.3|8.9|5.5$ & -\\
    Wav2Vec 2.0 (WV2) & 353.7 & 36.3 & 35.1 & 36.2 & $22.3|8.3|5.6$ & -\\
    Wav2Vec 2.0 Robust (WV2R) & 353.7 & 36.3 & 31.2 & 34.2 & $20.8|7.5|5.9$ & -\\
    WavLM (WL) & 351.7 & 36.3 & \textbf{24.9} & \textbf{27.6} & $15.6|6.3|5.6$ & -\\
    \hline
    WavLM + Fbank & 351.8 & 36.3 & 25.2 & 27.7 & $15.9|6.2|5.6$ & 58.5$+$41.5\\
    WavLM + Data2Vec & 665.1 & 36.4 & 25.0 & 27.2 & $15.8|\textbf{5.8}|5.6$ & 36.7$+$63.3\\
    WavLM + Wav2Vec 2.0 & 669.3 & 36.4 & 24.8 & 27.1 & $15.5|6.4|5.2$ & 58.9$+$41.1\\
    WavLM + Wav2Vec 2.0 Robust & 669.3 & 36.4 & 24.7 & 27.0 & $15.8|\textbf{5.8}|5.4$ & 59.0$+$41.0\\
    WavLM + HuBERT & 668.5 & 36.4 & \textbf{24.4} & \textbf{26.5} & $\textbf{15.2}|6.2|5.2$ & 53.4$+$46.6\\
    \hline
    WL + HB + WV2 & 986.0 & 36.5 & 24.7 & 27.0 & $15.6|6.0|5.4$ & 40.4$+$32.6$+$27.0\\
    WL + HB + WV2R & 986.0 & 36.5 & 24.8 & 27.0 & $15.5|6.3|\textbf{5.1}$ & 40.0$+$32.1$+$27.9\\
    WL + HB + D2V & 981.9 & 36.5 & 24.8 & 26.9 & $15.6|\textbf{5.8}|5.4$ & 27.7$+$23.5$+$48.8\\
    \bottomrule
  \end{tabular}
}
\label{tab:combine}
\end{table}
%%%%%%%%%%%%%%%%%%%%%%%%%%%%
% wavlm with different SSLR
In this part, we present results from various combination of features from different SSL models. Table~\ref{tab:combine} presents performance on the FSC Phase-4 corpus of leading SSL models previously shown to perform well on the SUPERB benchmark~\cite{yang2021superb}. Notably, WavLM achieves the lowest WER among individual models, demonstrating its resilience to noisy multi-speaker naturalistic audio.

We expanded the use of WavLM with each of Fbank, Data2Vec, Wav2Vec 2.0 Robust, Wav2Vec 2.0, and HuBERT models using a linear projection method for feature fusion. Despite Wav2Vec 2.0 Robust yielding the second-best WER individually, combining it with WavLM resulted in a less favorable outcome. Among all tested combinations, WavLM + HuBERT exhibits the best performance, achieving an absolute WER reduction of 1.1\% on Eval set compared to WavLM alone. This improvement is statistically significant with $p<0.001$, as verified using the MAPSSWE test \cite{gillick1989some}. We also evaluated three-SSL combinations, but none outperformed WavLM + HuBERT on the Eval set. While adding Wav2Vec 2.0 Robust to WavLM + HuBERT further lowers insertion error rate ($I=5.1$), this benefit is unfortunately offset by higher substitutions and deletions. 

To quantify the contribution of each representation, we analyzed the Frobenius norms of the learnable weight sub-matrices within the linear input projection layer (i.e., pre-encoder in Fig.~\ref{fig:frontend}), as shown in the last column of Table~\ref{tab:combine}. We observed that the best-performing pair, WavLM + HuBERT, exhibits a nearly balanced weight distribution (53.4\% vs. 46.6\%), suggesting that the downstream model effectively leverages complementary acoustic information from both representations.

In contrast, other combinations exhibited less effective weight distributions or domination by sub-optimal features. For instance, in WavLM + Data2Vec, the projection weights are heavily skewed toward Data2Vec (63.3\%), effectively suppressing the robust WavLM features (36.7\%) and degrading performance to 27.2\% WER. Furthermore, extending the fusion to three models (e.g., WL + HB + WV2R) results in significant weight redistribution. In this case, the contributions of the primary models (WavLM and HuBERT) drop to 40.0\% and 32.1\% respectively to accommodate the third representation. This redistribution yields no performance gain over the 2-model baselines (e.g., WavLM + Wav2Vec 2.0 Robust) and degrades performance compared to the optimal WavLM + HuBERT pair, implying that the third stream does not provide sufficient unique information to justify the dilution of the primary features.

% In contrast, WavLM + HuBERT reduces both substitutions and insertions relative to WavLM, producing the greatest net WER improvement (27.6\% $\rightarrow$ 26.5\%).
% However, HuBERT’s effect is \emph{combination-dependent}: when added on top of WavLM + Wav2Vec 2.0 (WL + HB + WV2 vs.\ WavLM + Wav2Vec 2.0), deletions fall but substitutions and insertions increase slightly, leading to only a marginal overall WER gain. Similarly, WL + HB + D2V achieves the lowest deletion error rate ($D=5.8$) among three-SSL variants, but increases substitution and insertion errors, resulting in worse performance than using WavLM + HuBERT. In summary, WavLM + HuBERT achieves the most favorable balance of substitution, deletion, and insertion error rates among all tested fusions.

\subsubsection{Phoneme Error Analysis}
\label{sec:phoneme_analysis}
\begin{table}[ht]
\centering
\caption{Phoneme class error totals for different SSL models on the Eval set of FSC Phase-4 corpus, based exclusively on phoneme alignments within word substitution errors. The HB, W2V2R, and WL stands for HuBERT, Wav2Vec 2.0 Robust, and WavLM. For the list of phonemes in each class, please see Appendix~A.}
\scalebox{0.88}{
\begin{tabular}{l|ccccccc}
\toprule
\textbf{SSL Models} & \textbf{Vowel} & \textbf{Stop} & \textbf{Fricative} & \textbf{Nasal} & \textbf{Liquid} & \textbf{Glide} & \textbf{Affricate} \\
\midrule
HB               & 45,678 & 26,212 & 17,954 & 12,712 & 8,907 & 5,927 & 1,543 \\
W2V2R            & 41,875 & 23,826 & 16,292 & 11,764 & 8,305 & 5,509 & 1,434 \\
WL               & 30,519 & 17,296 & 11,704 & 8,509  & 6,094 & 4,054 & 1,048 \\
WL + HB          & \textbf{29,641} & \textbf{16,818} & \textbf{11,289} & \textbf{8,047} & \textbf{5,720} & \textbf{3,885} & \textbf{975} \\
WL + W2V2R       & 31,266 & 17,581 & 11,952 & 8,571  & 5,984 & 4,052 & 1,021 \\
\bottomrule
\end{tabular}
}
\label{tab:phoneme_class_errors}
\end{table}
To gain further insight into the source of recognition errors and the benefits of SSL feature fusion, we conduct a phoneme-level analysis on the Eval set of the FSC Phase-4 corpus, focusing on the best-performing systems identified in Table~\ref{tab:combine}. Table~\ref{tab:phoneme_class_errors} presents the total number of phoneme errors for each major phoneme class (for the full list of phonemes in each class, see Appendix~A), based exclusively on phoneme alignments only within word substitution errors.

Among individual SSL models, WavLM exhibits the lowest error counts across all phoneme classes. The WavLM + HuBERT combination further reduces phoneme errors across all classes. In particular, the largest relative reductions compared to WavLM alone occur for affricates (-7.0\%), liquids (-6.1\%), and nasals (-5.4\%). On the other hand, WavLM + Wav2Vec 2.0 Robust only slightly improves the liquids, glides, and affricates while degrading all the others. 

Overall, the phoneme-level analysis shows that improvements in WER observed for WavLM + HuBERT (Sec.~\ref{sec:combine_ssl}) are supported by both broad, and consistent cross-class error reductions.

\subsubsection{Functional vs. Content Word Analysis}
\label{func_cont}
\begin{table}[ht]
\centering
\caption{Functional and content word error breakdown (substitution, deletion, insertion) on the Eval set of the FSC Phase-4 corpus. The HB, W2V2R, and WL stands for HuBERT, Wav2Vec 2.0 Robust, and WavLM. For the list of functional words, please see Appendix~B.}
\scalebox{0.85}{
\begin{tabular}{l|ccc|c||ccc|c}
\toprule
\textbf{SSL Models} & \multicolumn{4}{c||}{\textbf{Functional Words}} & \multicolumn{4}{c}{\textbf{Content Words}} \\
 & \textbf{Sub} & \textbf{Del} & \textbf{Ins} & \textbf{Sum} & \textbf{Sub} & \textbf{Del} & \textbf{Ins} & \textbf{Sum} \\
\midrule
HB         & 15,719 & 7,538 & 4,463 & 27,720 & 23,156 & 7,925 & 5,071 & 36,152 \\
W2V2R      & 14,518 & 6,404 & 4,763 & 25,685 & 21,677 & 6,681 & 5,528 & 33,886 \\
WL         & 10,282 & 5,345 & 4,378 & 20,005 & 16,888 & 5,713 & 5,358 & 27,959 \\
WL + HB    & \textbf{9,987} & 5,143 & \textbf{4,020} & \textbf{19,150} & \textbf{16,399} & 5,562 & \textbf{4,957} & \textbf{26,918} \\
WL + W2V2R & 10,546 & \textbf{4,963} & 4,122 & 19,631 & 16,902 & \textbf{5,225} & 5,298 & 27,425 \\
\bottomrule
\end{tabular}
}
\label{tab:func_cont}
\end{table}
Table~\ref{tab:func_cont} compares the error distributions for functional words (e.g., so-called ``stop'' words with limited information content; for the complete list, see Appendix~B) and content words across the evaluated SSL models from Table~\ref{tab:phoneme_class_errors}. WavLM + HuBERT yields consistent reductions for both categories, with a 4.3\% relative decrease in functional word errors, and 3.7\% in content word errors compared to WavLM alone. The largest gain is in functional word insertions, reduced by 8.2\%. In contrast, while WavLM + Wav2Vec~2.0 Robust achieves the lowest deletion errors for both categories, increases in substitution errors offset these gains, resulting in smaller overall improvements than with WavLM + HuBERT.

Although functional words contribute fewer absolute errors than content words, their misrecognition can disproportionately affect sentence structure and grammatical coherence. The improvements from WavLM + HuBERT are therefore valuable not only for lowering overall WER, but also for preserving the syntactic integrity of the text content under diverse noisy conditions seen in Apollo communications.

\subsection{Effect of Layer Selection in SSL Feature Fusion}
\begin{table}[th]
\centering
\caption{WER (\%) comparison using different layer selections from SSL models. "Top-$k$" denotes the weighted-sum of the $k$ highest-weighted SSL layers. The WL and HB stands for WavLM and HubERT.}
\scalebox{0.9}{
\begin{tabular}{l|c|c|c}
\toprule
\textbf{SSL Models} & \textbf{Layer Setting} & \textbf{Dev}$(\downarrow)$ & \textbf{Eval}$(\downarrow)$ \\
\midrule
WL & all & \textbf{24.9} & \textbf{27.6} \\
WL (Top-1) & L24 & 26.4 & 28.8 \\
WL (Top-3) & L21,22,24 & 25.0 & 28.1 \\
\midrule
WL + HB & all + all & \textbf{24.4} & \textbf{26.5} \\
WL + HB (Top-1) & L22 + L24 & 25.1 & 27.8 \\
WL + HB (Top-3) & L21,22,24 + L0,1,24 & 25.3 & 27.5 \\
\bottomrule
\end{tabular}
}
\label{tab:layer_selection}
\end{table}
Prior work has shown that lower layers of SSL models tend to encode more basic acoustic information, while upper layers capture more abstract linguistic or semantic characteristics \cite{hsu2021hubert, pasad2023comparative, ashihara2024self}. It has also been reported that using only the highest-weighted layers or empirically best performing layers can sometimes outperform the weighted-sum of all layers \cite{chiu2024learnable}. 

To investigate this, we compared different layer selection strategies for WavLM alone and for the WavLM + HuBERT fusion system as shown in Table~\ref{tab:layer_selection}. The highest-weighted layers were determined by inspecting the learned layer-combination weights in the fusion experiments in Table~\ref{tab:combine}. For example, HuBERT alone selects layers \{10, 12, 24\}, while in the fusion setup, HuBERT’s top-3 layers shift to \{0, 1, 24\}, suggesting that fusion benefits from incorporating lower-layer acoustic cues from HuBERT alongside higher-layer abstractions from WavLM.

Overall, Table~\ref{tab:layer_selection} shows that in both solo and fusion cases, the weighted-sum over all layers achieves the lowest WER. Accordingly, we adopt the weighted-sum of all layers for all subsequent experiments. Nevertheless, the shift in HuBERT’s preferred layers in fusion underscores that cross-model combinations can alter the relative utility of different layers, and that lower layers may contribute more when complementary information is available from another SSL model.

\subsection{Fusion Method Comparison}
% Fusion method comparison
Next, we explore fusion strategies for robust ASR on the FSC Phase-4 corpus. We evaluate various fusion methods, including our proposed DCA, and present their performance in Table~\ref{tab:fusion}. We first establish the WavLM as a single-SSL reference, which achieves a WER of 27.6\% on the Eval set. Incorporating features from HuBERT through weighted-sum fusion yields a modest improvement to 26.8\% WER. This method applies a trainable weighted combination of SSL features, but its simplicity limits potential gains. Linear projection and co-attention \cite{berrebbi2022combining} fusion both reduce WER to 26.5\%. Linear projection integrates features by projecting and concatenating them through a learnable transformation layer, while co-attention dynamically attends to relevant features across the two SSL models. However, their similar performance suggests that neither approach fully captures the complementary nature of SSL features in this challenging task.

Introducing Feature Refinement Loss (FRL) to the linear projection method imposes explicit constraints that refine the fused representation, yielding a further improvement to 26.4\% WER. Note that, although FRL proves effective when applied to HuBERT + Wav2Vec 2.0 (as shown in Table~\ref{tab:tune_refine_loss}), its impact is notably diminished in the WavLM + HuBERT setting here. This is likely because HuBERT and WavLM share the almost same architecture and training objective (i.e., masked prediction of pseudo labels), resulting in more similar feature representations and reducing the potential benefit of decorrelation. Comparing these results, the FRL regularizer provides a marginal gain (0.1\%) over the linear projection. In contrast, unlocking deeper feature interactions requires a more robust structural fusion approach.

%%% E-branchformer updated %%%
\begin{table}[t]
\centering
\caption{WER (\%) on FSC Phase-4 corpus when using alternate fusion methods. Note that FRL and DCA stands for Feature Refinement Loss and Deep Cross-Attention. The third column reports \textit{total} number of model parameters in millions (M), including frozen SSL models, and the fourth column shows number of \textit{trainable} parameters. E-Branchformer is used in these table results.}
\scalebox{0.8}{
  \begin{tabular}{l|c|c|c|c|c}
    \toprule
    \textbf{SSL Model} & \textbf{Fusion Method} & \textbf{Total(M)} & \textbf{Trainable(M)} & \textbf{Dev}$(\downarrow)$ & \textbf{Eval}$(\downarrow)$\\
    \hline
    WavLM & - & 351.7 & 36.3 & 24.9 & 27.6\\
    \hline
    WavLM + HuBERT & Weighted-Sum & 668.5 & 36.4 & 24.8 & 26.8\\
    WavLM + HuBERT & Co-Attention \cite{berrebbi2022combining} & 668.6 & 36.5 & 24.3 & 26.5\\
    WavLM + HuBERT & Linear Projection & 668.5 & 36.4 & 24.4 & 26.5\\
    WavLM + HuBERT & + FRL & 668.5 & 36.4 & 24.3 & 26.4\\
    WavLM + HuBERT & Linear Projection$^+$ & 675.8 & 43.7 & 24.1 & 26.3\\
    %WavLM + HuBERT & linear projection & 44.18 & 24.0 & 25.9\\ %% 16 encoders
    %WavLM + W2V2 Robust & DCA & 44.05 & \textbf{23.7} & 25.8\\
    WavLM + HuBERT & DCA & 676.1 & 44.1 & \textbf{23.7} & \textbf{25.7}\\
    % \hline
    % Wav2Vec 2.0 + HuBERT & Linear Projection &&\\
    % Wav2Vec 2.0 + HuBERT & + FRL &&\\
    % Wav2Vec 2.0 + HuBERT & DCA &&\\
    \bottomrule
  \end{tabular}
}
\label{tab:fusion}
\end{table}
%%%%%%%%%%%%%%%%%%%%%%%%%%%%

To address this limitation, we propose DCA, which is designed to better exploit feature complementarity even between closely related models. Our proposed DCA achieves the best result, with a WER of 25.7\% on the Eval set, representing a 1.1\% absolute improvement over the weighted-sum fusion. The DCA method achieves a statistically significant improvement over all other fusion methods, including Linear Projection$^+$, with $p < 0.001$ as measured by the MAPSSWE test \cite{gillick1989some}. These results suggest that DCA effectively captures nuanced complementary information between SSL model features, leveraging deep contextual interactions to yield better integration.

Furthermore, DCA increases the trainable parameters to 44.1 million, compared to approximately 36.4 million for other methods. To ensure this performance gain stems from the fusion mechanism rather than increased model capacity, we evaluated a scaled baseline, denoted as Linear Projection$^+$ in Table~\ref{tab:fusion}. This variant employs two linear layers with a hidden size of 3328 and GELU activation \cite{hendrycks2016gaussian}, matching DCA with 43.7 million trainable parameters. While Linear Projection$^+$ improves over the standard linear projection (26.3\% vs 26.5\%), it still underperforms DCA (25.7\%), confirming the validity of the cross-attention mechanism. However, we acknowledge that DCA entails higher computational complexity than simple projection methods. Additionally, the resulting improvement, while statistically significant ($p < 0.001$) and consistent across corpora, represents a modest 4.1\% relative gain (1.1\% absolute) over the weighted sum baseline. This indicates that DCA is a specialized strategy best suited for extracting complementary cues in challenging acoustic environments where simpler fusion methods saturate.

\subsection{WER Analysis on FSC Phase-4}
To understand where the WER improvement comes from in our solution, we show the WER on the Eval set of FSC Phase-4 corpus for each mission in Table~\ref{tab:wer_missions}. The table provides a detailed analysis of WER across Apollo missions, highlighting variations between seen/unseen channel conditions. For Apollo-8 (A8), the unseen PAO channel achieves a lower WER of 21.4\% compared to 26.6\% for seen channels, suggesting the model performs effectively in less complex and structured dialogues. This counterintuitive result may be attributed to the characteristics of the PAO channel, which often resembles a radio broadcast. PAO speech is typically more formal, slower-paced, and well-structured compared to the spontaneous, technical, and sometimes overlapping speech in mission control or onboard crew communications. These factors make the PAO channel easier for the model to transcribe accurately, despite it being unseen during training. Conversely, Apollo-11 (A11) shows a significant WER increase from seen conditions (23.0\%) to the unseen OPSPRO channel (30.0\%), indicating challenges associated with unfamiliar technical communication. Similarly, Apollo-13 (A13) exhibits a notable rise in WER, from 23.0\% in seen conditions to 31.1\% for the unseen Capsule Communicator (CAPCOM) channel, which may stem from the complexity of ground-to-space communication of CAPCOM. Across missions, a broader comparison reveals that Apollo-11 and Apollo-13 exhibit slightly higher overall WERs (25.7\% and 26.3\%, respectively) than A8 (25.5\%). This suggests that missions involving critical channels like OPSPRO and CAPCOM pose greater challenges. Overall, these findings highlight the necessity for ASR systems to handle diverse and unfamiliar communication to maintain accuracy.
\begin{table}[t]
    \centering
    \caption{The DCA system's WER (\%) on the Eval set of FSC Phase-4 corpus for each mission and channel under seen/unseen condition, with an overall WER of 25.7\%. The seen channels (NTWK, EECOM, GNC, FD, and MOCR) are those included in the training set, while A8 and A13 represent unseen missions. For details, please refer to Sec.~\ref{FSC_intro}.}
    \begin{tabular}{l|c|c}
        \toprule
        \textbf{Mission\_channel} & \textbf{WER}$(\downarrow)$ & \textbf{WER$(\downarrow)$ by Mission} \\ \hline
        A8\_seen & 26.6 & \multirow{2}{*}{25.5} \\
        A8\_unseen (PAO) & 21.4 & \\
        \midrule
        A11\_seen & 23.0 & \multirow{2}{*}{25.7} \\
        A11\_unseen (OPSPRO) & 30.0 & \\
        \midrule
        A13\_seen & 23.0 & \multirow{2}{*}{26.3} \\
        A13\_unseen (CAPCOM) & 31.1 & \\ \bottomrule
    \end{tabular}
    \label{tab:wer_missions}
\end{table}
\begin{figure*}[!t]
    \centering
    \subfloat[Dev set]{
    \includegraphics[width=0.8\linewidth]{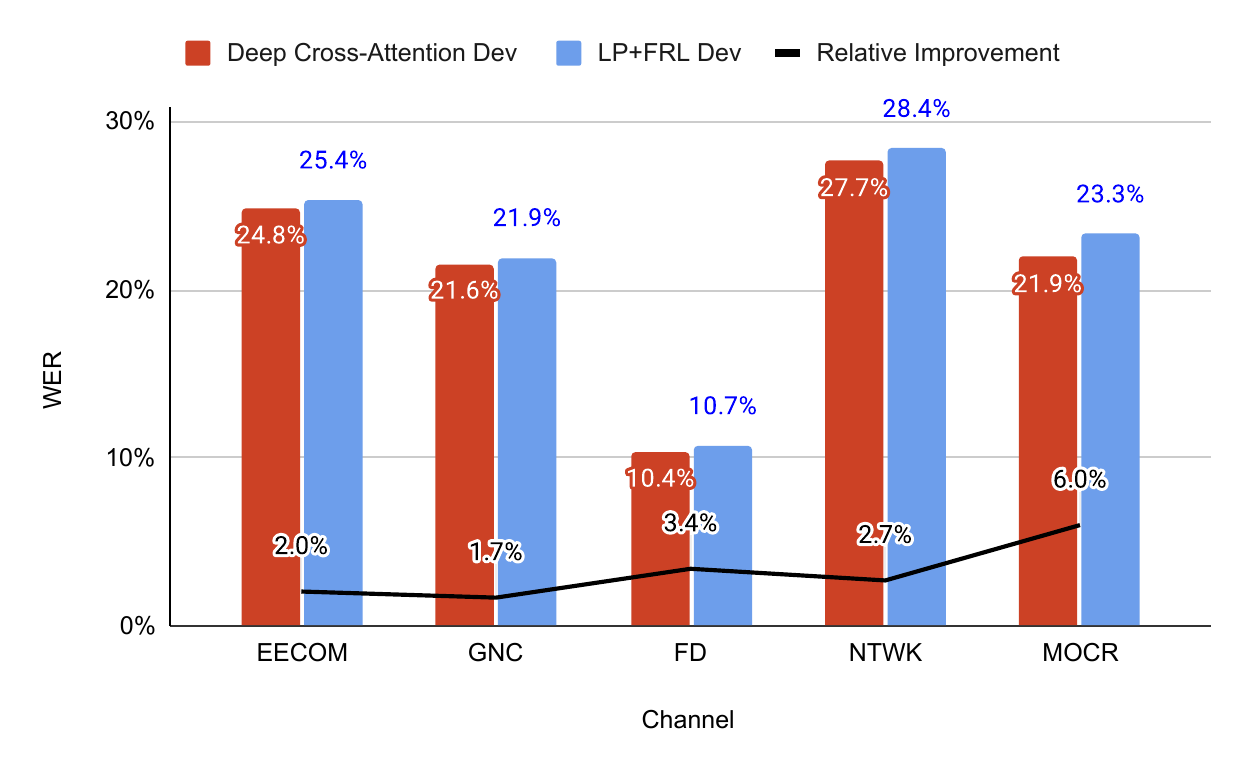}
    \label{fig:p4_dev}} \\%
    %\hspace*{\fill}
    \subfloat[Eval set]{
    \includegraphics[width=0.8\linewidth]{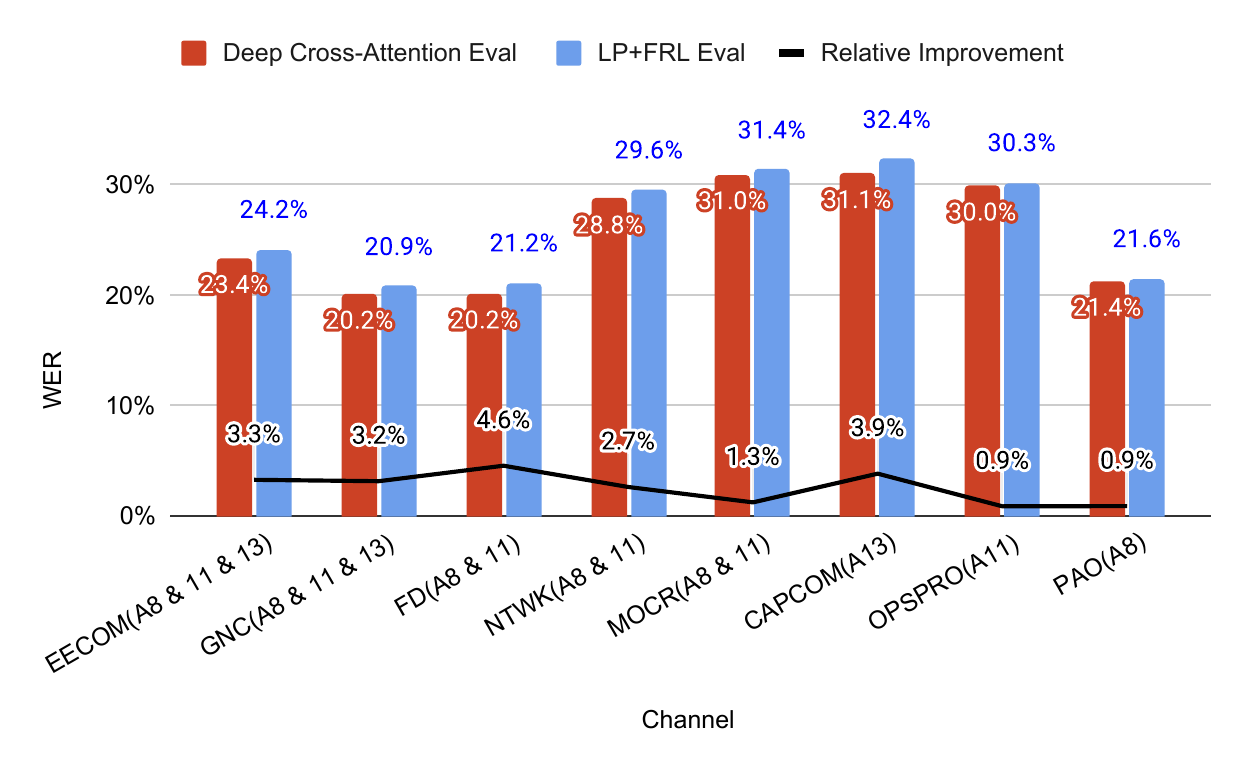}
    \label{fig:p4_eval}}
    \caption{Per-channel analysis of FSC Phase-4 corpus. WER (\%) shown for proposed Deep Cross-Attention (DCA) and linear projection + Feature Refinement Loss (LP+FRL). The relative WER benefits for proposed DCA over LP+FRL is shown for each channel. Note that the label "A" within the parenthesis denotes Apollo, e.g., A8 refers to Apollo-8.}
    \label{fig:per_analysis}
\end{figure*}

Additionally, we present a detailed per-channel WER analysis for the Dev and Eval sets of the FSC Phase-4 corpus (Fig.~\ref{fig:per_analysis}). Here, we compare the proposed DCA fusion method and the linear projection (LP) + FRL fusion method in Table~\ref{tab:fusion}. In the Dev set, the DCA system consistently outperforms the LP+FRL system across all channels. Notably, the MOCR channel benefits most significantly from DCA, with a relative improvement of 6.0\% over LP+FRL. The FD channel shows the lowest WER of 10.4\%, suggesting more manageable communication contexts. In the Eval set, the CAPCOM channel displays a high WER of 31.1\%, likely due to its core communications between Earth and Moon (e.g. NASA mission control and Astronauts in space/on the moon).
%Among all channels, the FD channel achieves the lowest WER and the MOCR channel benefits much more from the proposed method (see Fig.~\ref{fig:per_analysis}). On the other hand, the CAPCOM channel in Fig.~\ref{fig:p4_eval} has the highest WER, which might be due to its core communications between Earth and Moon (e.g. NASA mission control and Astronauts in space/on the moon).

% Since the Dev set contains only the original 5 selected channels of Apollo 11, we can ignore the effect of unseen channels and unseen missions  
% \begin{figure}[!t]
%   \begin{minipage}[t]{0.5\textwidth}
%     \centering
%     \includegraphics[width=\linewidth]{FSCp4_dev.pdf}
%     \caption{Dev set}
%     \label{fig:p4_dev}
%   \end{minipage}%
%   \begin{minipage}[t]{0.5\textwidth}
%     \centering
%     \includegraphics[width=\linewidth]{FSCp4_eval.pdf}
%     \caption{Eval set}
%     \label{fig:p4_eval}
%   \end{minipage}
%   \caption{Per-channel analysis of FSC Phase-4 corpus. The LP stands for the linear projection fusion with 52.63 million parameters. The label "A" in the channel name denotes Apollo, for instance, A8 refers to Apollo 8.}
%   \label{fig:per_analysis}
% \end{figure}

% \begin{table}[h]
%     \centering
%     \begin{tabular}{c|c|c|c}
%     \toprule
%     \textbf{Channel} & \textbf{Dev} & \textbf{Eval} & \textbf{Relative Improvement}
%     % \midrule
%     EECOM & 25.84 & 
%     \end{tabular}
%     \caption{Per-channel analysis on FSC Phase-4 corpus.}
%     \label{tab:per_channel}
% \end{table}

\begin{table}[t]
\centering
\caption{WER (\%) results of CHiME-6 dataset. Note that LP, FRL, and DCA stand for linear projection, Feature Refinement Loss, and Deep Cross-Attention. The third column reports \textit{total} number of model parameters in millions (M), including frozen SSL models, and the forth column shows number of \textit{trainable} parameters. We also use $\lambda=0.1$ for FRL. E-Branchformer is used in these table results.}
\scalebox{0.77}{
\begin{threeparttable}
  \begin{tabular}{l|c|c|c|c|c c}
    \toprule
    \textbf{SSL Model} & \textbf{Fusion Method} & \textbf{$\epsilon$} & \textbf{Total(M)} & \textbf{Trainable(M)} & \textbf{Dev}$(\downarrow)$ & \textbf{Eval}$(\downarrow)$\\
    \hline
    WavLM & - & - & 352.1 & 36.7 & 45.4 & 50.0 \\
    %HuBERT & - & - & 36.65 & 108.6 & 109.9 \\
    \hline
    WavLM + HuBERT & LP & - & 668.9 & 36.8 & 46.2 & 49.6\\ %p=0.038 to baseline
    WavLM + HuBERT & LP + FRL & 0.6 & 668.9 & 36.8 & 45.3 & 49.3 \\ %p=0.126 to LP
    WavLM + HuBERT & LP + FRL & 0.4 & 668.9 & 36.8 & 45.9 & 49.3  \\
    WavLM + HuBERT & LP + FRL & 0.2 & 668.9 & 36.8 & 46.0 & 49.6 \\
    WavLM + HuBERT & LP + FRL & 0.0 & 668.9 & 36.8 & 48.4 & 51.0 \\
    %\hline
    WavLM + HuBERT & Co-Attention \cite{berrebbi2022combining} & - & 668.9 & 36.9 & 54.0 & 57.4\\
    WavLM + HuBERT & DCA & - & 676.5 & 44.4 & \textbf{43.0} & \textbf{47.5}\\ %p<0.001
    \bottomrule
  \end{tabular}
\end{threeparttable}
} % end of scalebox
\label{tab:chime}
\end{table}
\subsection{CHiME-6 Result}
% CHiME-6 result
To further demonstrate the effectiveness of our proposed DCA fusion, we conducted experiments on the CHiME-6 corpus. Results are summarized in Table~\ref{tab:chime}. We choose WavLM as a single-SSL reference, which achieves a WER of 50.0\% on the Eval set. Combining WavLM with HuBERT using linear projection fusion improves WER slightly to 49.6\%. The addition of Feature Refinement Loss to the linear projection method is also evaluated under different correlation constraints ($\epsilon$). With $\epsilon = 0.6$, the linear projection + Feature Refinement Loss method achieves a WER of 49.3\% on the Eval set, outperforming linear projection alone. However, decreasing $\epsilon$ leads to degradation in performance, with WER rising to 51.0\% at $\epsilon = 0.0$. These findings align with observations in Sec.~\ref{FRL_result}, reinforcing that Feature Refinement Loss is most effective under moderate correlation constraints.

Surprisingly, co-attention fusion performs worse than both linear projection and the single-SSL WavLM model, with WER increasing significantly to 57.4\%. This suggests that co-attention struggles to effectively model intricate dependencies between features in the highly noisy and multi-speaker environment of CHiME-6. In contrast, our proposed DCA method achieves the best performance, significantly reducing WER to 47.5\% on the Eval set. This represents a statistically significant improvement ($p<0.001$) over the linear projection baseline and all other fusion methods. DCA's ability to effectively capture and leverage complementary information between SSL model features demonstrates its robustness, even in challenging acoustic conditions.

% \label{ablation}
% model size change
% \textbf{[To Dr. Hansen]} I will need to run the experiments using the proposed deep cross-attention and feature refinement loss when lowering model size. The purpose is to see if the methods are robust with model size.

\section{Conclusions}
\label{conclusion}
This work advances the study of self-supervised learning (SSL) feature fusion for automatic speech recognition (ASR) in naturalistic, noisy, and multi-speaker environments. In particular, we investigated Feature Refinement Loss by exploring its hyperparameters, experimenting with different maximum value of correlation between the extracted features. Our findings suggest that $\lambda=0.1$ and $\epsilon=0.6$ to be the optimal setting for the Fearless Steps Challenge (FSC) Phase-4 corpus. A better SSL model, WavLM, was also used in our study. We compared the WavLM with the top performing SSL models on the SUPERB benchmark \cite{yang2021superb} and choose the best combination that represented WavLM and HuBERT, for our SSL feature fusion experiments. Detailed error analyses and layer selection strategies were also conducted for the fusion systems to better understand the sources of performance improvements. 

Previous proposed fusion methods were first tested on the FSC Phase-4 corpus. However, we discovered that these methods often struggled to fully capture the complementary nature of features from different SSL models, particularly in highly challenging ASR tasks with multi-speakers and changing noisy environments. Hence, a novel deep cross-attention (DCA) fusion was proposed to address this problem of effective feature fusion. Our experiments showed that our proposed method yielded consistent, statistically significant improvements compared to all other fusion methods. While DCA entails higher computational complexity than simple projection baselines, its ability to capture deep feature interactions proves essential for preventing saturation in highly noisy scenarios. In addition, we conducted the same experiments on the separate CHiME-6 corpus to further confirm the effectiveness of the proposed DCA fusion method for naturalistic and adverse scenario, and showed our solution outperformed all other methods.

Most importantly, we presented the first ASR study and analysis of the FSC Phase-4 corpus, representing one of the first large scale, massive naturalistic team communications community resource corpora for the speech/language community. Compared to the previous state-of-the-art model and the popular huge pre-trained ASR model Whisper, a model solely using the extracted features from WavLM achieved the best WER on the FSC Phase-2 corpus, which was used as our strong baseline comparison model. We then showed performance mismatch of WavLM model on FSC Phase-2 and Phase-4 to demonstrate the severe challenges of the FSC Phase-4 corpus. In our per-channel WER analysis, we found that Mission Operations Control Room and Flight Director channels achieved the most improvement from our proposed SSL feature fusion method. We also noted that the Capsule Communicator channel has the worst WER among all channels, confirming that the Earth-to-space communications is severely challenging for effective ASR models.

In all, we have presented results on FSC Phase-4 corpus to showcase the ability of ASR model advancements for unseen channels and unseen missions. We also further pushed the boundary of SSL feature fusion by proposing the DCA fusion method. For future work, we could explore ways to use SSL models for better WER, in order to create higher quality Fearless Steps community meta-data resources for various disciplines including speech/language technology, education, preservation/history/archiving, communication science, Psychology/Small Group Teams. As a next step, we also plan to apply our proposed solution to the full 150,000 hours of Apollo data for public distribution and community resource sharing.

%%% Appendix usage %%%%
%see appendix~\ref{sec:sample:appendix}.

\appendix
\label{appendix}
% appendix text here
\section*{Appendix A: Phoneme Classes Used in Sec.~\ref{sec:phoneme_analysis}}
For the phoneme class error analysis in Sec.~\ref{sec:phoneme_analysis}, we grouped CMU-style phonemes into the following categories:

\begin{itemize}
    \item \textbf{Vowels:} \texttt{\{AA, AE, AH, AO, AW, AY, EH, ER, EY, IH, IY, OW, OY, UH, UW\}}
    \item \textbf{Stops:} \texttt{\{B, D, G, K, P, T\}}
    \item \textbf{Fricatives:} \texttt{\{DH, F, S, SH, TH, V, Z, ZH\}}
    \item \textbf{Nasals:} \texttt{\{M, N, NG\}}
    \item \textbf{Affricates:} \texttt{\{CH, JH\}}
    \item \textbf{Liquids:} \texttt{\{L, R\}}
    \item \textbf{Glides:} \texttt{\{W, Y, HH\}}
\end{itemize}

\section*{Appendix B: List of Functional Words Used in Sec.~\ref{func_cont}}

The following is a categorized list of functional words used in functional and content word analysis:

\textbf{Determiners and Articles:}
\vspace{-0.3cm}
\begin{multicols}{3}
\noindent
a\\ an\\ the\\ this\\ that\\ these\\ those\\ each\\ every\\ all\\ some\\ any\\ no
\end{multicols}

% \vspace{0.5em}
\textbf{Coordinating Conjunctions:}
\vspace{-0.3cm}
\begin{multicols}{3}
\noindent
and\\ but\\ or\\ nor\\ so\\ for\\ yet
\end{multicols}

% \vspace{0.5em}
\textbf{Subordinating Conjunctions / Connectives:}
\vspace{-0.3cm}
\begin{multicols}{3}
\noindent
because\\ as\\ if\\ while\\ although\\ though\\ unless\\ until\\ since\\ once\\ when\\ whenever\\ before\\ after
\end{multicols}

% \vspace{0.5em}
\textbf{Prepositions:}
\vspace{-0.3cm}
\begin{multicols}{3}
\noindent
of\\ at\\ by\\ for\\ with\\ about\\ against\\ between\\ into\\ through\\ during\\ above\\ below\\ to\\ from\\ up\\ down\\ in\\ out\\ on\\ off\\ over\\ under\\ around\\ within\\ without
\end{multicols}

% \vspace{0.5em}
\textbf{Pronouns:}
\vspace{-0.3cm}
\begin{multicols}{3}
\noindent
i\\ you\\ he\\ she\\ it\\ we\\ they\\ me\\ him\\ her\\ us\\ them\\ my\\ your\\ his\\ its\\ our\\ their\\ mine\\ yours\\ hers\\ ours\\ theirs
\end{multicols}

% \vspace{0.5em}
\textbf{Modals and Auxiliaries:}
\vspace{-0.3cm}
\begin{multicols}{3}
\noindent
can\\ could\\ shall\\ should\\ will\\ would\\ may\\ might\\ must\\ do\\ does\\ did\\ have\\ has\\ had\\ am\\ is\\ are\\ was\\ were\\ be\\ being\\ been
\end{multicols}

% \vspace{0.5em}
\textbf{Common Adverbs:}
\vspace{-0.3cm}
\begin{multicols}{3}
\noindent
again\\ further\\ then\\ there\\ here\\ very\\ too\\ just\\ not\\ also\\ still
\end{multicols}

% \vspace{0.5em}
\textbf{WH-words:}
\vspace{-0.3cm}
\begin{multicols}{3}
\noindent
what\\ which\\ who\\ whom\\ whose\\ why\\ how\\ where\\ when
\end{multicols}

% \vspace{0.5em}
\textbf{Particles and Miscellaneous:}
\vspace{-0.3cm}
\begin{multicols}{3}
\noindent
than\\ only\\ own\\ such
\end{multicols}

 \bibliographystyle{elsarticle-num} 
 \bibliography{cas-refs}

@inproceedings{chen2021crossvit,
  title={Crossvit: Cross-attention multi-scale vision transformer for image classification},
  author={Chen, Chun-Fu Richard and Fan, Quanfu and Panda, Rameswar},
  booktitle={Proceedings of the IEEE/CVF international conference on computer vision},
  pages={357--366},
  year={2021}
}

@inproceedings{lin2022cat,
  title={Cat: Cross attention in vision transformer},
  author={Lin, Hezheng and Cheng, Xing and Wu, Xiangyu and Shen, Dong},
  booktitle={2022 IEEE international conference on multimedia and expo (ICME)},
  pages={1--6},
  year={2022},
  organization={IEEE}
}

@inproceedings{cai2024car,
  title={CAR-Transformer: Cross-Attention Reinforcement Transformer for Cross-Lingual Summarization},
  author={Cai, Yuang and Yuan, Yuyu},
  booktitle={Proceedings of the AAAI Conference on Artificial Intelligence},
  volume={38},
  pages={17718--17726},
  year={2024}
}

@article{xiong2018microsoft,
  title={{The Microsoft 2017 conversational speech recognition system}},
  author={Xiong, Wayne and Wu, Lingfeng and Alleva, Fil and Droppo, Jasha and Huang, Xuedong and Stolcke, Andreas},
  journal={{IEEE ICASSP-2018: Inter. Conf. on Acoustics, Speech and Signal Proc.}},
  pages={5934--5938},
  year={2018}
}

@article{watanabe2017hybrid,
  title={{Hybrid CTC/attention architecture for End-to-End speech recognition}},
  author={Watanabe, Shinji and Hori, Takaaki and Kim, Suyoun and Hershey, John R and Hayashi, Tomoki},
  journal={IEEE Journal of Selected Topics in Signal Processing},
  volume={11},
  number={8},
  pages={1240--1253},
  year={2017},
  publisher={IEEE}
}

@article{gorin400houston,
  title={{“This is Houston. Say again, please.” The Behavox system for the {Apollo-11 Fearless Steps Challenge (Phase II)}}},
  author={Gorin, Arseniy and Kulko, Daniil and Grima, Steven and Glasman, Alex},
  journal={ISCA Interspeech-2020},
  pages={2612--2616},
  year={2020},
}

@article{tuske2021limit,
  title={{On the limit of English conversational speech recognition}},
  author={T{\"u}ske, Zolt{\'a}n and Saon, George and Kingsbury, Brian},
  journal={ISCA Interspeech-2021},
  year={2021},
}

@article{kim2023branchformer,
  title={{E-Branchformer: Branchformer with Enhanced merging for speech recognition}},
  author={Kim, Kwangyoun and Wu, Felix and Peng, Yifan and Pan, Jing and Sridhar, Prashant and Han, Kyu J and Watanabe, Shinji},
  journal={SLT-23: IEEE Spoken Lang. Tech. Workshop},
  pages={84--91},
  year={2023},
  //organization={IEEE}
}

@article{radford2022robust,
  title={Robust speech recognition via large-scale weak supervision},
  author={Radford, Alec and Kim, Jong Wook and Xu, Tao and Brockman, Greg and McLeavey, Christine and Sutskever, Ilya},
  journal={Intern. Conference on Machine Learning},
  pages={28492--28518},
  year={2023},
  organization={PMLR},
}

@article{vaswani2017attention,
  title={Attention is all you need},
  author={Vaswani, Ashish and Shazeer, Noam and Parmar, Niki and Uszkoreit, Jakob and Jones, Llion and Gomez, Aidan N and Kaiser, {\L}ukasz and Polosukhin, Illia},
  journal={Advances in Neural Info. Proc. Systems},
  volume={30},
  pages={5998-6008},
  year={2017}
}

@article{gulati2020conformer,
  title={{Conformer: Convolution-augmented Transformer for Speech Recognition}},
  author={Gulati, Anmol and Qin, James and Chiu, Chung-Cheng and Parmar, Niki and Zhang, Yu and Yu, Jiahui and Han, Wei and Wang, Shibo and Zhang, Zhengdong and Wu, Yonghui and others},
  journal={Proc. Interspeech 2020},
  pages={5036--5040},
  year={2020}
}

@article{chi-etal-2022-xlm,
    title = "{XLM}-{E}: Cross-lingual Language Model Pre-training via {ELECTRA}",
    author = "Chi, Zewen  and
      Huang, Shaohan  and
      Dong, Li  and
      Ma, Shuming  and
      Zheng, Bo  and
      Singhal, Saksham  and
      Bajaj, Payal  and
      Song, Xia  and
      Mao, Xian-Ling  and
      Huang, Heyan  and
      Wei, Furu",
    journal = "Annual Meeting of Assoc. for Comp. Ling.",
    year = "2022",
    doi = "10.18653/v1/2022.acl-long.427",
    pages = "6170--6182",
}

@article{hsu2021hubert,
  title={Hubert: Self-supervised speech representation learning by masked prediction of hidden units},
  author={Hsu, Wei-Ning and Bolte, Benjamin and Tsai, Yao-Hung Hubert and Lakhotia, Kushal and Salakhutdinov, Ruslan and Mohamed, Abdelrahman},
  journal={IEEE/ACM Trans. on Audio, Speech, and Language Proc.},
  volume={29},
  pages={3451--3460},
  year={2021},
  publisher={IEEE}
}

@article{baevski2020wav2vec,
  title={{Wav2Vec 2.0: A framework for self-supervised learning of speech representations}},
  author={Baevski, Alexei and Zhou, Yuhao and Mohamed, Abdelrahman and Auli, Michael},
  journal={Advances in Neural Information Processing Systems},
  volume={33},
  pages={12449--12460},
  year={2020}
}

@article{chen2022wavlm,
  title={{WavLM: Large-scale self-supervised pre-training for full stack speech processing}},
  author={Chen, Sanyuan and Wang, Chengyi and Chen, Zhengyang and Wu, Yu and Liu, Shujie and Chen, Zhuo and Li, Jinyu and Kanda, Naoyuki and Yoshioka, Takuya and Xiao, Xiong and others},
  journal={IEEE Journal of Selected Topics in Signal Processing},
  volume={16},
  number={6},
  pages={1505--1518},
  year={2022},
  publisher={IEEE}
}

@article{oord2018representation,
  title={Representation learning with contrastive predictive coding},
  author={Oord, Aaron van den and Li, Yazhe and Vinyals, Oriol},
  journal={Proc. of NIPS},
  year={2018}
}

@article{chang2021exploration,
  title={An exploration of self-supervised pretrained representations for end-to-end speech recognition},
  author={Chang, Xuankai and Maekaku, Takashi and Guo, Pengcheng and Shi, Jing and Lu, Yen-Ju and Subramanian, Aswin Shanmugam and Wang, Tianzi and Yang, Shu-wen and Tsao, Yu and Lee, Hung-yi and others},
  journal={IEEE ASRU-2021: Automatic Speech Recog. and Understanding Workshop},
  pages={228--235},
  year={2021},
}

@article{mohamed2022self,
  title={Self-supervised speech representation learning: A review},
  author={Mohamed, Abdelrahman and Lee, Hung-yi and Borgholt, Lasse and Havtorn, Jakob D and Edin, Joakim and Igel, Christian and Kirchhoff, Katrin and Li, Shang-Wen and Livescu, Karen and Maal{\o}e, Lars and others},
  journal={IEEE Journal of Selected Topics in Signal Processing},
  volume={16},
  number={6},
  pages={1179--1210},
  year={2022},
  publisher={IEEE}
}

@article{Joglekar2021,
  author={Aditya Joglekar and Seyed Omid Sadjadi and Meena Chandra-Shekar and Christopher Cieri and John H.L. Hansen},
  title={{Fearless Steps Challenge Phase-3 (FSC P3): Advancing SLT for Unseen Channel and Mission Data Across NASA Apollo Audio}},
  year=2021,
  journal={ISCA Interspeech-2021},
  pages={986--990},
  doi={10.21437/Interspeech.2021-2011}
}

@article{joglekar2020fearless,
  title={{FEARLESS STEPS Challenge (FS-2): Supervised Learning with Massive Naturalistic Apollo Data}},
  author={Joglekar, Aditya and Hansen, John HL and Shekar, Meena Chandra and Sangwan, Abhijeet},
  journal={ISCA Interspeech-2020},
  pages={2617--2621},
  year={2020}
}

@article{Hansen2019,
  author={John H.L. Hansen and Aditya Joglekar and Meena Chandra Shekhar and Vinay Kothapally and Chengzhu Yu and Lakshmish Kaushik and Abhijeet Sangwan},
  title={{The 2019 Inaugural Fearless Steps Challenge: A Giant Leap for Naturalistic Audio}},
  year=2019,
  journal={ISCA Interspeech-2019},
  pages={1851--1855},
  doi={10.21437/Interspeech.2019-2301},
  url={http://dx.doi.org/10.21437/Interspeech.2019-2301}
}

@article{hansen2018fearless,
  title={Fearless {S}teps: Apollo-11 {C}orpus {A}dvancements for {S}peech {T}echnologies from {E}arth to the {M}oon},
  author={Hansen, John HL and Sangwan, Abhijeet and Joglekar, Aditya and Bulut, Ahmet Emin and Kaushik, Lakshmish and Yu, Chengzhu},
  journal={ISCA Interspeech-2018},
  pages={2758--2762},
  year={2018}
}

@inproceedings{hansen2024fearless,
  title={Fearless Steps Apollo: Team Communications Based Community Resource Development for Science, Technology, Education, and Historical Preservation},
  author={Hansen, John HL and Joglekar, Aditya and Shekar, Meena MC and Chen, Szu-Jui and Liu, Xi},
  booktitle={ICASSP 2024-2024 IEEE International Conference on Acoustics, Speech and Signal Processing (ICASSP)},
  pages={12816--12820},
  year={2024},
  organization={IEEE}
}

@article{panayotov2015librispeech,
  title={Librispeech: an asr corpus based on public domain audio books},
  author={Panayotov, Vassil and Chen, Guoguo and Povey, Daniel and Khudanpur, Sanjeev},
  journal={IEEE ICASSSP-2015: Intern. Conf. on Acoustics, Speech and Signal Proc.},
  pages={5206--5210},
  year={2015},
  organization={IEEE}
}

@article{watanabe2020chime,
  title={{CHiME-6 challenge: Tackling multispeaker speech recognition for unsegmented recordings}},
  author={Watanabe, Shinji and Mandel, Michael and Barker, Jon and Vincent, Emmanuel and Arora, Ashish and Chang, Xuankai and Khudanpur, Sanjeev and Manohar, Vimal and Povey, Daniel and Raj, Desh and others},
  journal={Workshop on Speech Processing in Everyday Environments (CHiME 2020)},
  pages={1--7},
  year={2020}
}

@inproceedings{boeddeker2018front,
  title={Front-end processing for the CHiME-5 dinner party scenario},
  author={Boeddeker, Christoph and Heitkaemper, Jens and Schmalenstroeer, Joerg and Drude, Lukas and Heymann, Jahn and Haeb-Umbach, Reinhold},
  booktitle={CHiME5 Workshop, Hyderabad, India},
  volume={1},
  year={2018}
}

@article{yang2021superb,
  title={Superb: Speech processing universal performance benchmark},
  author={Yang, Shu-wen and Chi, Po-Han and Chuang, Yung-Sung and Lai, Cheng-I Jeff and Lakhotia, Kushal and Lin, Yist Y and Liu, Andy T and Shi, Jiatong and Chang, Xuankai and Lin, Guan-Ting and others},
  journal={ISCA Interspeech-2021},
  pages={1194-1198},
  year={2021},
}

@article{arunkumar2022investigation,
  title={Investigation of Ensemble features of Self-Supervised Pretrained Models for Automatic Speech Recognition},
  author={Arunkumar, A and Sukhadia, Vrunda N and Umesh, Srinivasan},
  journal={ISCA Interspeech-2022},
  pages={5145-5149},
  year={2022}
}

@article{berrebbi2022combining,
  title={Combining Spectral and Self-Supervised Features for Low Resource Speech Recognition and Translation},
  author={Berrebbi, Dan and Shi, Jiatong and Yan, Brian and Lopez-Francisco, Osbel and Amith, Jonathan D and Watanabe, Shinji},
  journal={ISCA Interspeech-2022},
  pages={3533-3537},
  year={2022}
}

@article{chen2022fearless,
  title={{FeaRLESS: Feature Refinement Loss for Ensembling Self-Supervised Learning Features in Robust End-to-end Speech Recognition}},
  author={Chen, Szu-Jui and Xie, Jiamin and Hansen, John H. L.},
  journal={ISCA Interspeech-2022},
  year={2022}
}

@article{chen2021scenario,
  title={{Scenario Aware Speech Recognition: Advancements for Apollo Fearless Steps \& CHiME-4 Corpora}},
  author={Chen, Szu-Jui and Xia, Wei and Hansen, John H. L.},
  journal={IEEE ASRU-2021: Automatic Speech Recog. and Understanding Workshop},
  pages={289--295},
  year={2021},
  organization={IEEE}
}

@inproceedings{srivastava2024effuse,
  title={EFFUSE: Efficient Self-Supervised Feature Fusion for E2E ASR in Low Resource and Multilingual Scenarios},
  author={Srivastava, Tejes and Shi, Jiatong and Chen, William and Watanabe, Shinji},
  booktitle={Proc. Interspeech 2024},
  pages={3989--3993},
  year={2024}
}

@inproceedings{wang2024fusion,
  title={Fusion Of Discrete Representations and Self-Augmented Representations for Multilingual Automatic Speech Recognition},
  author={Wang, Shih-Heng and Shi, Jiatong and Huang, Chien-yu and Watanabe, Shinji and Lee, Hung-yi},
  booktitle={2024 IEEE Spoken Language Technology Workshop (SLT)},
  pages={247--254},
  year={2024},
  organization={IEEE}
}

@inproceedings{chiu2024learnable,
  title={Learnable Layer Selection and Model Fusion for Speech Self-Supervised Learning Models},
  author={Chiu, Sheng-Chieh and Wu, Chia-Hua and Hsieh, Jih-Kang and Tsao, Yu and Wang, Hsin-Min},
  booktitle={Proc. Interspeech 2024},
  pages={3914--3918},
  year={2024}
}

@article{watanabe2018espnet,
  title={ESPnet: End-to-End Speech Processing Toolkit},
  author={Watanabe, Shinji and Hori, Takaaki and Karita, Shigeki and Hayashi, Tomoki and Nishitoba, Jiro and Unno, Yuya and Soplin, Nelson-Enrique Yalta and Heymann, Jahn and Wiesner, Matthew and Chen, Nanxin and others},
  journal={Proc. Interspeech 2018},
  pages={2207--2211},
  year={2018}
}

@article{nguyen2020investigating,
  title={Investigating self-supervised pre-training for end-to-end speech translation},
  author={Nguyen, Ha and Bougares, Fethi and Tomashenko, Natalia and Est{\`e}ve, Yannick and Besacier, Laurent},
  journal={ISCA Interspeech-2020},
  year={2020}
}

@article{wu2020self,
  title={Self-supervised representations improve end-to-end speech translation},
  author={Wu, Anne and Wang, Changhan and Pino, Juan and Gu, Jiatao},
  journal={ISCA Interspeech-2020},
  pages={1491--1495},
  year={2020},
}

@article{yi2020applying,
  title={{Applying Wav2Vec 2.0 to speech recognition in various low-resource languages}},
  author={Yi, Cheng and Wang, Jianzhong and Cheng, Ning and Zhou, Shiyu and Xu, Bo},
  journal={arXiv preprint arXiv:2012.12121},
  year={2020}
}

@article{fan2020exploring,
  title={{Exploring Wav2Vec 2.0 on speaker verification and language identification}},
  author={Fan, Zhiyun and Li, Meng and Zhou, Shiyu and Xu, Bo},
  journal={ISCA Interspeech-2021},
  pages={1509-1513},
  year={2021}
}

@article{pepino2021emotion,
  title={{Emotion recognition from speech using Wav2Vec 2.0 embeddings}},
  author={Pepino, Leonardo and Riera, Pablo and Ferrer, Luciana},
  journal={ISCA Interspeech-2021},
  pages={3400--3404},
  year={2021}
}

@article{kingma2014adam,
  title={Adam: A method for stochastic optimization},
  author={Kingma, Diederik P and Ba, Jimmy},
  journal={International Conference on Learning Representations},
  year={2014}
}

@article{loshchilov2017decoupled,
  title={Decoupled weight decay regularization},
  author={Loshchilov, I},
  journal={arXiv preprint arXiv:1711.05101},
  year={2017}
}

@article{park2019specaugment,
  title={Specaugment: A simple data augmentation method for automatic speech recognition},
  author={Park, Daniel S and Chan, William and Zhang, Yu and Chiu, Chung-Cheng and Zoph, Barret and Cubuk, Ekin D and Le, Quoc V},
  journal={Proc. Annu. Conf. Int. Speech Commun. Assoc.},
  pages={2613–2617},
  year={2019}
}

@article{hendrycks2016gaussian,
  title={Gaussian error linear units (gelus)},
  author={Hendrycks, Dan and Gimpel, Kevin},
  journal={arXiv preprint arXiv:1606.08415},
  year={2016}
}

@inproceedings{gillick1989some,
  title={Some statistical issues in the comparison of speech recognition algorithms},
  author={Gillick, Laurence and Cox, Stephen J},
  booktitle={International Conference on Acoustics, Speech, and Signal Processing,},
  pages={532--535},
  year={1989},
  organization={IEEE}
}

@misc{fiscus2018sctk,
  author       = {Fiscus, Jonathan},
  title        = {{NIST SCTK Toolkit}},
  year         = {2018},
  howpublished = {\url{https://github.com/usnistgov/SCTK}},
  note         = {Online;}
}

@inproceedings{pasad2023comparative,
  title={Comparative layer-wise analysis of self-supervised speech models},
  author={Pasad, Ankita and Shi, Bowen and Livescu, Karen},
  booktitle={ICASSP 2023-2023 IEEE International Conference on Acoustics, Speech and Signal Processing (ICASSP)},
  pages={1--5},
  year={2023},
  organization={IEEE}
}

@inproceedings{ashihara2024self,
  title={What do self-supervised speech and speaker models learn? new findings from a cross model layer-wise analysis},
  author={Ashihara, Takanori and Delcroix, Marc and Moriya, Takafumi and Matsuura, Kohei and Asami, Taichi and Ijima, Yusuke},
  booktitle={ICASSP 2024-2024 IEEE International Conference on Acoustics, Speech and Signal Processing (ICASSP)},
  pages={10166--10170},
  year={2024},
  organization={IEEE}
}





% \end{thebibliography}
\end{document}